\documentclass[10pt, conference, twocolumn]{IEEEtran}
\usepackage{cite}
\usepackage{xspace}
\usepackage{xcolor}
\usepackage{graphicx}
\usepackage{amsmath}
\usepackage{paralist}
\usepackage{caption}
\usepackage{nicefrac, xfrac}
\usepackage{rotating}
\usepackage{blkarray}
\usepackage{wrapfig}
\usepackage{multirow}

\usepackage{url}

\newcommand{\relu}{ReLU\xspace{}}

\newcommand{\sat}{\texttt{SAT}}
\newcommand{\unsat}{\texttt{UNSAT}}
\newcommand{\error}{\texttt{ERROR}}
\newcommand{\timeout}{\texttt{TIMEOUT}}

\usepackage[ruled]{algorithm}
\usepackage[noend]{algorithmic}
\usepackage[inline]{enumitem}
\usepackage{varwidth}

\newcommand{\nn}{neural network}

\newcommand{\vq}{verification query}

\newcommand{\nnv}{neural network verifier}

\newcommand{\assign}[2]{{#1}_{w}(#2)}

\newcommand{\ourTool}{\textsc{DelBugV}} 

\newcommand{\mysubsection}[1]{\medskip\noindent\textbf{#1}}

\usepackage{tikz,ifthen,pgfplots}

\usetikzlibrary{arrows,trees,backgrounds,automata,shapes,decorations,plotmarks,fit,calc,positioning,shadows,chains}

\definecolor{color0}{RGB}{228,87,46}
\definecolor{color1}{RGB}{23,190,187}
\definecolor{color2}{RGB}{255,201,20}
\definecolor{color3}{RGB}{46,40,42}
\definecolor{color4}{RGB}{118,176,65}

\definecolor{color0}{RGB}{250,121,33}
\definecolor{color1}{RGB}{254,153,32}
\definecolor{color2}{RGB}{185,164,76}
\definecolor{color3}{RGB}{86,110,61}
\definecolor{color4}{RGB}{12,71,103}

\definecolor{color0}{RGB}{228,253,225}
\definecolor{color1}{RGB}{138,203,136}
\definecolor{color2}{RGB}{100,131,129}
\definecolor{color3}{RGB}{87,87,97}
\definecolor{color4}{RGB}{255,191,70}

\definecolor{color0}{RGB}{226,59,62}
\definecolor{color1}{RGB}{243,114,44}
\definecolor{color2}{RGB}{248,150,30}
\definecolor{color3}{RGB}{249,199,79}
\definecolor{color4}{RGB}{126,179,86}
\definecolor{color5}{RGB}{67,170,139}
\definecolor{color6}{RGB}{39,125,161}
\definecolor{color7}{RGB}{21,49,60}
\definecolor{color8}{RGB}{180,215,228}
\definecolor{color9}{RGB}{251, 217, 137}

\tikzstyle{every pin edge}=[<-,shorten <=1pt]
\tikzstyle{neuron}=[circle,fill=black!25,minimum size=17pt,inner sep=0pt]
\tikzstyle{input neuron}=[neuron, fill=green!50]
\tikzstyle{output neuron}=[neuron, fill=red!50]
\tikzstyle{hidden neuron}=[neuron, fill=blue!50]
\tikzstyle{annot} = [text width=4em, text centered]
\usetikzlibrary{calc}

\definecolor{nnedgecolor}{RGB}{90,90,90}
\tikzstyle{every pin edge}=[<-,shorten <=1pt]
\tikzstyle{every path}=[draw=color7!50]
\tikzstyle{neuron}=[circle,fill=black!25,minimum size=17pt,inner sep=0pt]
\tikzstyle{input neuron}=[neuron, fill=color4]
\tikzstyle{output neuron}=[neuron, fill=color0]
\tikzstyle{hidden neuron}=[neuron, fill=color6!80]
\tikzstyle{bias neuron}=[neuron, fill=color9!80]
\tikzstyle{assignment neuron}=[neuron, fill=none]
\tikzstyle{annot} = [text width=4em, text centered]
\tikzstyle{nnedge} = [-{stealth},shorten >=0.1cm, shorten <=0.05cm,line width=0.8pt,nnedgecolor]
\usetikzlibrary{calc}

\begin{document}
\title{DelBugV: Delta-Debugging Neural\\ Network Verifiers}

\author{
\IEEEauthorblockN{Raya Elsaleh and Guy Katz}
\IEEEauthorblockA{The Hebrew University of Jerusalem, Jerusalem, Israel
}}
 

\maketitle
\begin{abstract}
  Deep neural networks (DNNs) are becoming a key component in diverse
  systems across the board. However, despite their success, they often
  err miserably; and this has triggered significant interest in
  formally verifying them. Unfortunately, DNN verifiers are intricate
  tools, and are themselves susceptible to soundness bugs. Due to the
  complexity of DNN verifiers, as well as the sizes of the DNNs being
  verified, debugging such errors is a daunting task. Here, we present
  a novel tool, named \ourTool{}, that uses automated \emph{delta
    debugging} techniques on DNN verifiers. Given a malfunctioning DNN verifier
  and a correct verifier as a point of reference (or, in some cases,
  just a single, malfunctioning verifier), \ourTool{} can produce
  much simpler DNN verification instances that still trigger undesired
  behavior --- greatly facilitating the task of debugging the faulty
  verifier. Our tool is modular and extensible, and can easily be
  enhanced with additional network simplification methods and
  strategies. For evaluation purposes, we ran \ourTool{} on 4 DNN
  verification engines, which were observed to produce incorrect
  results at the 2021 neural network verification competition (VNN-COMP'21). We were able to simplify many of the verification
  queries that trigger these faulty behaviors, by as much as 99\%. We
  regard our work as a step towards the ultimate goal of producing
  reliable and trustworthy DNN-based software.
\end{abstract}

\section{Introduction}
Deep neural networks (DNNs)~\cite{Goodfellow-et-al-2016} are software
artifacts that are generated automatically, through the generalization
of a finite set of examples. These artifacts have been shown to outdo
manually crafted software in a variety of key domains, such as natural
language processing~\cite{goldberg2016primer, liu2019multi,
  guo2020gluoncv}, image recognition~\cite{wu2015deep,
  guo2020gluoncv}, protein folding~\cite{noe2020machine,
  hou2018deepsf}, and many others.  However, this impressive success
comes at a price: unlike traditional software, DNNs are opaque
artifacts, and are incomprehensible to humans. This poses a serious
challenge when it comes to certifying, modifying, extending, repairing
or reasoning about them~\cite{huang2017safety, katz2019marabou,
  43405}.

In an effort to address these issues, the formal methods community has
taken up an interest in \emph{DNN verification}~\cite{KaBaDiJuKo17,
  huang2017safety, PuTa10}: automated techniques that can determine
whether a DNN satisfies a prescribed specification, and provide a
counter-example if it does not. DNN verification technology has been
making great strides, and its applicability has been demonstrated in
various domains~\cite{KaBaDiJuKo17, GeMiDrTsChVe18, KoLoJaBl20,
  AmKaSc22, AmFrKaMaRe23, AmScKa21}.  In fact, this technology has progressed to
a point where DNN verifiers themselves have become quite complex, and
consequently error-prone; especially as they often perform delicate
arithmetic operations, which can also introduce bugs into the
verification process~\cite{KaBaDiJuKo17}. Thus, it is not surprising
that various bugs have been observed in these
tools~\cite{JiRi20b}. For example, in the VNN-COMP'21
competition~\cite{https://doi.org/10.48550/arxiv.2109.00498}, various
verifiers have been shown to disagree on the result of multiple
verification queries (each query is comprised of a neural network and a 
property to be checked), or produce incorrect counter-examples,
indicating the existence of bugs.  Moreover, many of these verifiers
are still being developed, with new and experimental features being
introduced --- potentially introducing new bugs as well.  An
inability to trust the results of DNN verifiers could undermine the
benefits of DNN verification technology, and clearly needs to be
addressed.

Here, we propose to mitigate this issue by adopting known techniques
from related fields (e.g., SMT solving~\cite{10.1145/1670412.1670413})
--- specifically, that of  \emph{delta
  debugging}. The idea is to leverage the fact that DNN verification
is at a point where many verification tools are available, and to allow
engineers to readily compare the results produced by their verification tool to
those produced by others, in order to identify and correct bugs. 
When a verification query that triggers some bug in 
a verifier is detected, we can initiate
an automated process that repeatedly and incrementally
\emph{simplifies} the verification query. After each
simplification step, we can check that the verifier in question still
disagrees with the remaining, \emph{oracle} verifiers, until reaching the simplest
verification query that we can find. If this final query is much
simpler than the original, it will be that much easier for engineers
to debug their tools, eventually improving their overall soundness.

We present a new tool, \ourTool{} (\textbf{Del}ta de\textbf{Bug}ging
Neural Network \textbf{V}erifiers),
that takes as input a 
verification query, a malfunctioning DNN verifier that errs
on the given verification query, and
an oracle DNN verifier. Within \ourTool{}, we 
implement a set of operations for simplifying the neural network of the given verification query into a network with fewer layers and fewer
neurons. We empirically design a strategy that applies these
operations sequentially in an order that produces much
simpler verification queries. In some cases, when the 
malfunctioning DNN verifier produces a
faulty counter-example, \ourTool{} can run in \emph{single solver}
mode -- without an oracle verifier, where the query is repeatedly 
simplified as long as the malfunctioning DNN verifier
continues to produce incorrect counter-examples.

For evaluation, we tested \ourTool{} on 
4 DNN verifiers ``suspected'' of errors, per the results of
 VNN-COMP'21~\cite{https://doi.org/10.48550/arxiv.2109.00498}:
 Marabou~\cite{katz2019marabou, WuZeKaBa22, OsBaKa22},
NNV~\cite{10.1007/978-3-030-53288-8_2, inbook23, tran2019star,
  https://doi.org/10.48550/arxiv.2004.05519, XiTrJo18},
NeuralVerification.jl(NV.jl)~\cite{LiArLaBaKo20}, 
and nnenum~\cite{bak2020execution, 10.1007/978-3-030-53288-8_2, tran2019star, bak2021nnenum}.
We ran \ourTool{} on queries where pairs of these verifiers disagreed. 
Our evaluation demonstrates that
\ourTool{} could reduce the size of the error-triggering queries by an
average of $96.8\%$, and by as much as $99\%$ in some cases, resulting in very simple neural networks. We believe that
these results highlight the significant potential of our tool and
approach.

The rest of the paper is organized as follows. In
Sec.~\ref{sec:background} we provide the necessary background on DNNs
and their verification. Next, in Sec.~\ref{sec:tool} we describe the
design of \ourTool{}, focusing on its algorithm and network
simplification methods and the strategy we use to apply those methods. The implementation and evaluation of \ourTool{} is
discussed in Sec.~\ref{sec:evaluation}. This is followed by a
discussion of related work in Sec.~\ref{sec:relatedWork}, and we
conclude in Sec.~\ref{sec:conclusion}.
\section{Background}
\label{sec:background}
\mysubsection{Neural Networks.}  A \emph{\nn{}} is a directed acyclic graph in
which the nodes, called neurons, are organized in layers
$l^0, l^1, \ldots, l^n$.  $l^0$ is called the input layer, $l^n$ the
output layer, and layers $l^1, \ldots, l^{n-1}$ are called  hidden
layers.  Each hidden layer has an associated non-linear
\emph{activation function}.  In feed-forward networks, which are our
subject matter here, neurons in layer $l^i$ have edges connecting them only to
neurons in the next layer, layer $l^{i+1}$.

Each neuron in the network (except the ones in the input layer)  
has a bias value, and each edge has a
weight. The biases and weights belonging to neurons in layer $l^i$ are
organized into a vector $B^i$ and a matrix $W^i$, respectively.  
The $j,j'$-th entry of $W^i$ is the weight assigned to the edge
out-going from the $j'$-th neuron in layer $l^{i-1}$ and entering the $j$-th
neuron in layer $l^{i}$.  For a \emph{fully connected} layer, $W^i$ is
a full matrix; whereas for a \emph{convolutional} layer, $W^i$ is very
sparse, and has a specific structure (discussed later).

An input to \nn{} $\mathcal{N}$ is a vector $I$ of values of the neurons in the input layer, and it produces an output vector $\mathcal{N}(I)$ which
is the values of the neurons in the output layer. We denote the values
of neurons in layer $l^i$, prior to applying the activation function, 
by \(\mathcal{N}^{l^i}(I)\); and the values after applying the
activation function by \(\mathcal{N}^{a^i}(I)\). The values of the neurons are
evaluated according to the rules:
\begin{align*}
  \mathcal{N}^{l^0}(I) = I, & \qquad
  \mathcal{N}^{l^i}(I) = W^{i} \mathcal{N}^{a^{i-1}}(I) + B^i, \\
  & \mathcal{N}^{a^i}(I) = Act^i(\mathcal{N}^{l^{i}}(I))
\end{align*} 
where $Act^i$ is the activation function associated with layer $l_i$.

We define the size of a \nn{} to be the total number of neurons in the
graph (including the neurons in the input and output layers) and denote
it by $|\mathcal{N}|$. 
The automated training
(i.e., selection of weights and biases)
of neural networks is beyond
our scope here; see, e.g.,~\cite{Goodfellow-et-al-2016}.

\begin{figure}[ht]
  \centering
  \scalebox{0.85}{
  \def\WSsep{2cm}
  \def\BSsep{-0.0cm}
  \def\IOWSsep{1.5cm}
  \def\relusep{1.4cm}
  \begin{tikzpicture}[shorten >=1pt,->,draw=black, node distance=\layersep,font=\footnotesize, no arrow/.style={-,every loop/.append style={-}}]

    \draw [no arrow,draw=black,decorate,very thick,decoration={brace,amplitude=5pt,raise=0.2cm}]
    (-0.75cm,-0.75) --  (0.75cm,-0.75) node[midway,yshift=0.7cm]{Input layer $l^0$};
    \draw [no arrow,draw=black,decorate,very thick,decoration={brace,amplitude=5pt,raise=0.2cm}]
    (\IOWSsep + 0 * \relusep + 0 * \WSsep + 0.3cm + \BSsep-0.5cm,1) --  (\IOWSsep + 1 * \relusep + 0 * \WSsep + 0.3cm+0.5cm,1) node[midway,yshift=0.7cm]{First hidden layer $l^1$};
    \draw [no arrow,draw=black,decorate,very thick,decoration={brace,amplitude=5pt,raise=0.2cm}]
    (\IOWSsep + 1 * \relusep + 1 * \WSsep + 0.3cm + \BSsep-0.5cm,1) --  (\IOWSsep + 2 * \relusep + 1 * \WSsep + 0.3cm+0.5cm,1) node[midway,yshift=0.7cm]{Second hidden layer $l^2$};
    \draw [no arrow,draw=black,decorate,very thick,decoration={brace,amplitude=5pt,raise=0.2cm}]
    (2 * \IOWSsep + 2 * \relusep + 1 * \WSsep + 0.3cm-0.75cm,-0.75) --  (2 * \IOWSsep + 2 * \relusep + 1 * \WSsep + 0.3cm+0.75cm,-0.75) node[midway,yshift=0.7cm]{Output layer $l^3$};

    \node[input neuron] (I-0) at (0,-2) {$x$};
    \node[assignment neuron] () at (0, -0.5 -2) {$5$};

    \node[hidden neuron] (H-1-0) at (\IOWSsep + 0 * \relusep + 0 * \WSsep + 0.3cm,0) {$l^1_0$};
    \node[hidden neuron] (H-1-1) at (\IOWSsep + 0 * \relusep + 0 * \WSsep + 0.3cm,-2) {$l^1_1$};
    \node[hidden neuron] (H-1-2) at (\IOWSsep + 0 * \relusep + 0 * \WSsep + 0.3cm,-4) {$l^1_2$};

    \node[bias neuron] (B-1-0) at (\IOWSsep + 0 * \relusep + 0 * \WSsep + 0.3cm + \BSsep,0.75) {$+10$};
    \node[bias neuron] (B-1-0) at (\IOWSsep + 0 * \relusep + 0 * \WSsep + 0.3cm + \BSsep,0.75 -2) {$-2.5$};
    \node[bias neuron] (B-1-0) at (\IOWSsep + 0 * \relusep + 0 * \WSsep + 0.3cm + \BSsep,0.75 -4) {$+7$};

    \node[assignment neuron] () at (\IOWSsep + 0 * \relusep + 0 * \WSsep + 0.3cm,-0.5) {$-15$};
    \node[assignment neuron] () at (\IOWSsep + 0 * \relusep + 0 * \WSsep + 0.3cm,-0.5 -2) {$-5$};
    \node[assignment neuron] () at (\IOWSsep + 0 * \relusep + 0 * \WSsep + 0.3cm,-0.5 -4) {$2$};

    \node[hidden neuron] (R-1-0) at (\IOWSsep + 1 * \relusep + 0 * \WSsep + 0.3cm,0) {$r^1_0$};
    \node[hidden neuron] (R-1-1) at (\IOWSsep + 1 * \relusep + 0 * \WSsep + 0.3cm,-2) {$r^1_1$};
    \node[hidden neuron] (R-1-2) at (\IOWSsep + 1 * \relusep + 0 * \WSsep + 0.3cm,-4) {$r^1_2$};

    \node[assignment neuron] () at (\IOWSsep + 1 * \relusep + 0 * \WSsep + 0.3cm,-0.5) {$0$};
    \node[assignment neuron] () at (\IOWSsep + 1 * \relusep + 0 * \WSsep + 0.3cm,-0.5 -2) {$0$};
    \node[assignment neuron] () at (\IOWSsep + 1 * \relusep + 0 * \WSsep + 0.3cm,-0.5 -4) {$2$};

    \node[hidden neuron] (H-2-0) at (\IOWSsep + 1 * \relusep + 1 * \WSsep + 0.3cm,0) {$l^2_0$};
    \node[hidden neuron] (H-2-1) at (\IOWSsep + 1 * \relusep + 1 * \WSsep + 0.3cm,-2) {$l^2_1$};
    \node[hidden neuron] (H-2-2) at (\IOWSsep + 1 * \relusep + 1 * \WSsep + 0.3cm,-4) {$l^2_2$};

    \node[bias neuron] (B-2-0) at (\IOWSsep + 1 * \relusep + 1 * \WSsep + 0.3cm + \BSsep,0.75) {$+8$};
    \node[bias neuron] (B-2-1) at (\IOWSsep + 1 * \relusep + 1 * \WSsep + 0.3cm + \BSsep,0.75-2) {$+2$};

    \node[assignment neuron] () at (\IOWSsep + 1 * \relusep + 1 * \WSsep + 0.3cm,-0.5) {$4$};
    \node[assignment neuron] () at (\IOWSsep + 1 * \relusep + 1 * \WSsep + 0.3cm,-0.5 -2) {$2$};
    \node[assignment neuron] () at (\IOWSsep + 1 * \relusep + 1 * \WSsep + 0.3cm,-0.5 -4) {$-2$};

    \node[hidden neuron] (R-2-0) at (\IOWSsep + 2 * \relusep + 1 * \WSsep + 0.3cm,0) {$r^2_0$};
    \node[hidden neuron] (R-2-1) at (\IOWSsep + 2 * \relusep + 1 * \WSsep + 0.3cm,-2) {$r^2_1$};
    \node[hidden neuron] (R-2-2) at (\IOWSsep + 2 * \relusep + 1 * \WSsep + 0.3cm,-4) {$r^2_2$};

    \node[assignment neuron] () at (\IOWSsep + 2 * \relusep + 1 * \WSsep + 0.3cm,-0.5) {$4$};
    \node[assignment neuron] () at (\IOWSsep + 2 * \relusep + 1 * \WSsep + 0.3cm,-0.5 -2) {$2$};
    \node[assignment neuron] () at (\IOWSsep + 2 * \relusep + 1 * \WSsep + 0.3cm,-0.5 -4) {$0$};
    
    \node[output neuron] (O-0) at (2 * \IOWSsep + 2 * \relusep + 1 * \WSsep + 0.3cm,-2) {$y$};
    \node[assignment neuron] () at (2 * \IOWSsep + 2 * \relusep + 1 * \WSsep + 0.3cm,-0.5 -2) {$5$};    

    \draw[nnedge] (I-0) --node[above,left,pos=0.37] {$-5$\;} (H-1-0);
    \draw[nnedge] (I-0) --node[above,pos=0.25] {\;$-0.5$} (H-1-1);
    \draw[nnedge] (I-0) --node[above,pos=0.25] {\;$-1$} (H-1-2);
    
    \draw[nnedge] (H-1-0) --node[above] {$\relu$} (R-1-0);
    \draw[nnedge] (H-1-1) --node[above] {$\relu$} (R-1-1);
    \draw[nnedge] (H-1-2) --node[above] {$\relu$} (R-1-2);
    
    \draw[nnedge] (R-1-0) --node[above,pos=0.2] {$0.8$} (H-2-0);
    \draw[nnedge] (R-1-0) --node[above,right,pos=0.1] {$2$} (H-2-2);
    
    \draw[nnedge] (R-1-1) --node[above,left,pos=0.25] {$-1$} (H-2-0);
    \draw[nnedge] (R-1-1) --node[above,pos=0.25] {$0.5$} (H-2-1);
    \draw[nnedge] (R-1-1) --node[above,right,pos=0.05] {$0.5$} (H-2-2);
    
    \draw[nnedge] (R-1-2) --node[above,left,pos=0.15] {$-2$} (H-2-0);
    \draw[nnedge] (R-1-2) --node[above,pos=0.2] {$-1$} (H-2-2);
    
    \draw[nnedge] (H-2-0) --node[above] {$\relu$} (R-2-0);
    \draw[nnedge] (H-2-1) --node[above] {$\relu$} (R-2-1);
    \draw[nnedge] (H-2-2) --node[above] {$\relu$} (R-2-2);
    
    \draw[nnedge] (R-2-0) --node[above,right,pos=0.5] {\;$0.25$} (O-0);
    \draw[nnedge] (R-2-1) --node[above,pos=0.55] {$2$} (O-0);
    \draw[nnedge] (R-2-2) --node[below,right,pos=0.45] {$\;0.5$} (O-0);
    
  \end{tikzpicture}
  }
  \captionsetup{justification=centering}
  \caption{$\mathcal{N}_e$ An example of a neural network with \relu{} activation functions.}
  
  \label{fig:MainExample}
\end{figure}
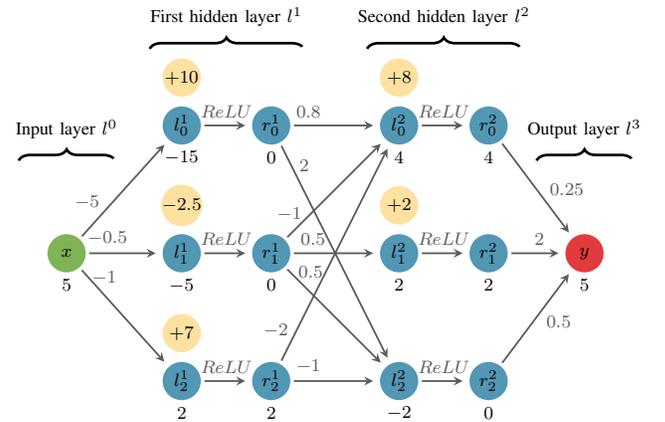

Fig.~\ref{fig:MainExample} depicts a \nn{}, \(\mathcal{N}_e\), with a
single input, a single output, and 2 hidden layers with 3 neurons in each. It uses the
\relu{} activation function, $\relu(x)=\max(0,x)$. The bias of each neuron is listed
above it, and weights are listed over the edges (zero values  are omitted). In matrix representation, the weights and
biases are:
\begin{align*}
  W^{1}=\begin{bmatrix}
  -5\\
  -0.5\\
  -1
  \end{bmatrix}& ,B^1 = \begin{bmatrix}10\\ -2.5\\ 7\end{bmatrix},
  W^{2} = \begin{bmatrix}
  0.8 & -1 & -2\\
  0 & 0.5 & 0\\
  2 & 0.5 & -1
  \end{bmatrix} ,\\ B^2 &= \begin{bmatrix}8\\ 2\\ 0\end{bmatrix} ,
  W^{3}= \begin{bmatrix}0.25 \\ 2 \\ 0.5\end{bmatrix}^T,
  B^3=\begin{bmatrix}0\end{bmatrix}
\end{align*}
\(\mathcal{N}_e\) is of size 8  (every $l^i_j$ and $r^i_j$  pair in the 
figure are counted as one neuron; we split them only for visualization
purposes), and has 4 layers.
The figure also demonstrates an evaluation of the network,  for the input
$x=5$. The assignment of each node is listed below it; and we can see
that the produced output in this case is $y=5$.

\mysubsection{Convolutional Neural Networks.}  A \emph{convolutional
  \nn{}} is a \nn{} with one or more convolutional layers (typically,
these are the first layers of the network).  The parameters of a
convolutional layer include the height $h$ and width $w$ of images in
the input; the kernel size $k$; the stride size $s$; the padding size
$p$; the input channels $c_i$; the output channels $c_o$; the kernel
weights $W$, given as a tensor of dimensions
$(c_o \times c_i \times k \times k)$; and the biases, $B$, organized
in an array of length $c_o$.  We assume for simplicity that the kernel
size, padding size, and stride size are equal along all axes, although
this is not a limitation of our approach.

The convolutional layer filters its input, which is a $(c_i \times k \times k)$-dimensional matrix,  using the above parameters and  outputs a multidimensional matrix which represents feature maps.
For additional information on how a convolutional layer computes its output, see~\cite{Goodfellow-et-al-2016}. Note that convolutional layers are comprised strictly of linear operations.

\mysubsection{Neural Network Verification.}  A \emph{property}
$\mathcal{P}$ is a set of constraints on the inputs and outputs of the
\nn{}.  These constraints give rise to an input region
$I(\mathcal{P})$ and an output region $O(\mathcal{P})$.  Verifying
$\mathcal{P}$, with respect to some neural network, 
entails determining whether there exists an input in
$I(\mathcal{P})$ that the \nn{} maps to an output in $O(\mathcal{P})$
(the \sat{} case), or not (the \unsat{} case).  Typically,
$\mathcal{P}$ is specified so that $O(\mathcal{P})$ represents
\emph{undesirable} behavior, and so an \unsat{} result indicates that
the system is correct.
$\mathcal{P}_e = (5\leq x\leq 10) \wedge (5\leq y\leq 10) $ is an example
of a property of $\mathcal{N}_e$ in Fig.~\ref{fig:MainExample}.


A \emph{\nnv{}} takes in a \vq{} (a neural network and a property) 
and attempts to automatically verify it. When successful, it  returns a
\sat{} or \unsat{} answer; otherwise, it can return $ \error{}$, or $\timeout{}$.
When a \nnv{} returns \sat{}, it also returns an input that proves the
satisfiability of the query.  Given a verifier \(\mathcal{V}\) and a
\vq{} \(Q=(\mathcal{N}, \mathcal{P})\), we denote by
\(\mathcal{V}(Q)\in \{\sat{}, \unsat{}, \error{}, \timeout{}\}\) the
answer of \(\mathcal{V}\) on $Q$. If
\(\mathcal{V}(Q) = \sat{}\), we denote by
\(\assign{\mathcal{V}}{Q} \in I(\mathcal{P})\) the satisfying
assignment (the witness) returned by the verifier. 

Continuing with our running example, given a sound 
\nnv{} $\mathcal{V}_e$ and 
the \vq{} $Q_e =(\mathcal{N}_e, \mathcal{P}_e)$, 
\(\mathcal{V}_e(Q_e) = \sat{}\) 
and a valid witness is \(\assign{\left(\mathcal{V}_e\right)}{Q_e} = (5)\),
since \( \mathcal{N}_e((5)) = (5) \in O(\mathcal{P}_e) \).

Neural network verification is complex, both theoretically and
practically~\cite{KaBaDiJuKo17}; and modern tools apply
sophisticated techniques to verify large networks~\cite{Al21}. These
techniques are typically theoretically sound, but implementation bugs
can cause verifiers to produce incorrect results. These bugs
are easier to track and correct if the problem manifests for queries
with small networks.

In a situation where two verifiers disagree on the satisfiability of a
given query, at least one of them must answer \sat{} and provide a
satisfying assignment. We evaluate the
neural network on that assignment, and determine whether it indeed
satisfies the property at hand. If so, we conclude that the other
verifier, which returned \unsat{}, is faulty; otherwise, if the
satisfying assignment is incorrect, we determine that the verifier
that answered \sat{} is faulty. The remaining verifier then takes the
role of the oracle verifier.

\section{\ourTool{}: Delta-Debugging Verification Queries}
\label{sec:tool}

\subsection{General Flow}
Applying \emph{delta-debugging} techniques means automatically 
simplifying an input $x$ that triggers a bug in the system into 
a simpler input, $x'$, that also triggers a bug~\cite{NiPrBa22}. 
$x'$ can often trigger the bug faster, thus
reducing overall debugging time; and also trigger fewer code lines that are
unrelated to the bug,  allowing engineers to more easily identify
its root cause.
In our setting, given a \vq{} \(Q=(\mathcal{N}, \mathcal{P})\) that
triggers a bug in a \nnv{}, we seek to generate another query \(Q'=(\mathcal{N}', \mathcal{P})\), with a much smaller
(simplified) neural network: $|\mathcal{N}'|<|\mathcal{N}|$. The motivation for focusing on the neural
network, and not on the verification conditions, is that common
verification conditions are typically already quite simple~\cite{nnvcomp20},
whereas neural network sizes have a crucial effect on verifier
performance~\cite{KaBaDiJuKo17}.

The general delta debugging framework that our tool  follows
appears as Alg.~\ref{alg:generalFlow}. The inputs to the process are a
faulty verifier $\mathcal{V}$, an oracle verifier $\mathcal{V}_O$, and
a verification query $Q=(\mathcal{N},\mathcal{P})$. The algorithm
maintains a candidate result neural network $\mathcal{N}_r$ that
triggers a bug in $\mathcal{V}$ and make it produce an incorrect answer,
and whose size is iteratively decreased. In each iteration, the
algorithm invokes Alg.~\ref{alg:simplify} to attempt simplifying $\mathcal{N}_r$.
The process terminates when
 Alg.~\ref{alg:simplify} states that it cannot simplify
 $\mathcal{N}_r$ any further, or
when a timeout limit is exceeded. 
Finally, it returns the verification query with the smallest $\mathcal{N}_r$ it  achieved.


\begin{algorithm}
  \caption{\textit{Reduce Verification Query}} 
  \begin{algorithmic}[1]
    \renewcommand{\algorithmicrequire}{\textbf{Input:}}
    \renewcommand{\algorithmicensure}{\textbf{Output:}}
    \REQUIRE
    $\mathcal{V}$, $\mathcal{V}_O$, $Q=(\mathcal{N}, \mathcal{P})$
    {\color{blue}\\// \textit{Faulty Verifier, Oracle Verifier, Verification query}}
    \ENSURE $Q_r$ {\color{blue}//\textit{ A simplified 
        query}}
    \STATE $\mathcal{N}_r\leftarrow \mathcal{N}$
    \STATE progressMade $\leftarrow$ True
    \WHILE { noTimeout() $\wedge$ progressMade }
    \STATE $\mathcal{N}_r\leftarrow \mathcal{N}$
    \STATE progressMade, $\mathcal{N}\leftarrow$ Simplify($\mathcal{V}, \mathcal{V}_O, Q$) \label{line:lineSimplify}
    \ENDWHILE
    \RETURN $(\mathcal{N}_r, \mathcal{P})$
  \end{algorithmic}
  \label{alg:generalFlow}
\end{algorithm}
\medskip

Alg.~\ref{alg:simplify} takes in the same arguments as
Alg.~\ref{alg:generalFlow}, and its goal is to perform one successful
simplification step on $\mathcal{N}$, from a pool of potential
steps. The algorithm heuristically chooses a sequence of
simplification steps to attempt (Line~\ref{line:pickAttempts}), and
then performs them, one by one, until one is successful. We propose
several simplification steps in Sec.~\ref{ssec:methods}.
Specifying the order according to which theses simplification steps are attempted
(Line~\ref{line:pickAttempts}) 
is key, and
different strategies may result in different
simplified networks ---
we propose one such strategy in Sec.~\ref{ssec:methods}.


\begin{algorithm}
  \caption{\textit{Simplify}} 
  \begin{algorithmic}[1]
    \renewcommand{\algorithmicrequire}{\textbf{Input:}}
    \renewcommand{\algorithmicensure}{\textbf{Output:}}
    \REQUIRE
    $\mathcal{V}$, $\mathcal{V}_O$, $Q=(\mathcal{N}, \mathcal{P})$
    {\color{blue}\\
    // \textit{Faulty Verifier, Oracle Verifier, Verification query}}
    \ENSURE True/False, $Q_r$ {\color{blue}//\textit{ Whether the
        query was 
        simplified, and the simplified 
        query}}
    
    \STATE Attempts = $\left(M_0, M_1,\ldots\right)\leftarrow$ \\ \qquad \qquad \qquad \quad attemptsBySimplificationStrategy$(\mathcal{N})$ \label{line:pickAttempts}
    \WHILE { Attempts $\neq \emptyset$ }
    \STATE $M_i \leftarrow $Attempts$.pop()$
    \STATE $\mathcal{N}_r\leftarrow M_i (\mathcal{N})$
    \IF {
      successSimplification($\mathcal{V},\mathcal{V}_O,(\mathcal{N}_r,
      \mathcal{P})$) } \label{line:checkIfSuccessful}
    \RETURN True, $\mathcal{N}_r$
    \ENDIF
    \ENDWHILE
    \RETURN False, $\mathcal{N}$
  \end{algorithmic}
  \label{alg:simplify}
\end{algorithm}
\medskip

Line~\ref{line:checkIfSuccessful} of
Alg.~\ref{alg:simplify} invokes
Alg.~\ref{alg:successSimplification} to check whether the simplification
step attempted succeeded or not.  To do so, Alg.~\ref{alg:successSimplification} first checks whether $\mathcal{V}$ answers \sat{}, but
returns an incorrect counter-example. If so, this
candidate should clearly be kept. Otherwise, the algorithm checks whether $\mathcal{V}$ and
$\mathcal{V}_O$ both answer \unsat{} or \sat{}, but disagree; if so,
it returns True. In all other cases, i.e. where one of the verifiers times
out, or when there is no basis for comparison (one of the verifiers returned an error), the algorithm returns False, and
an alternative simplification step in Alg.~\ref{alg:simplify} is attempted.

\begin{algorithm}
  \caption{\textit{successSimplification}} 
  \begin{algorithmic}[1]
    \renewcommand{\algorithmicrequire}{\textbf{Input:}}
    \renewcommand{\algorithmicensure}{\textbf{Output:}}
    \REQUIRE
    $\mathcal{V}$, $\mathcal{V}_O$, $Q=(\mathcal{N}, \mathcal{P})$
    {\color{blue}\\
    // \textit{Faulty Verifier, Oracle Verifier, Verification query}}
    \ENSURE True/False {\color{blue}//\textit{ Was the query successfully 
    simplified?}}

      \IF{
      $\mathcal{V}(\mathcal{N}, \mathcal{P})=\sat \wedge \mathcal{V}_W(Q)\notin I(\mathcal{P})$ }
      \STATE \textbf{return} True 
      \ENDIF
      \IF{
      $  \mathcal{V}(\mathcal{N}, \mathcal{P})=\sat \wedge \mathcal{N}(\mathcal{V}_W(Q))\notin O(\mathcal{P})$   }
        \STATE \textbf{return} True 
      \ENDIF
      \IF{$\mathcal{V}(\mathcal{N}, \mathcal{P}),
      \mathcal{V}_O(\mathcal{N}, \mathcal{P})\in \{\sat,\unsat\}$
      \\ \qquad \qquad  \qquad \qquad  $\wedge \ 
      \mathcal{V}(\mathcal{N}, \mathcal{P})\neq
      \mathcal{V}_O(\mathcal{N}, \mathcal{P})$ }
        \STATE \textbf{return} True 
      \ENDIF
    \RETURN False
  \end{algorithmic}
  \label{alg:successSimplification}
\end{algorithm}

One possible risk when using Alg.~\ref{alg:generalFlow} is a ``flip''
between the two verifiers. This can happen when initially,
$\mathcal{V}_O$ produces a correct answer and $\mathcal{V}$ does not;
but after a simplification step, $\mathcal{V}$ starts producing the correct
answer and $\mathcal{V}_O$ starts producing an incorrect answer. 
This situation is
unlikely: the simplification steps we propose later make local
modifications to the network, and  are consequently far more likely to
continue to trigger the same bug in $\mathcal{V}$ than to trigger a
new one in $\mathcal{V}_O$. Still, this concern can be mitigated even
further by using multiple oracle verifiers, and ensuring that they all
agree amongst themselves while $\mathcal{V}$ dissents. 

\mysubsection{Single Verifier Mode.}
Our approach could also be applied to delta-debug a single verifier that
returns incorrect satisfying assignments, without using an
oracle.
As we explain in Sec.~\ref{ssec:methods}, the simplification methods we apply require the returned satisfying assignment from either the faulty or the oracle verifier, thus, if the faulty verifier returns an incorrect satisfying assignment for the query at hand,  we can drop the oracle verifier.
This is achieved by removing the last ``if''
condition from Alg.~\ref{alg:successSimplification} and removing the oracle verifier $\mathcal{V}_O$ from the inputs.

\subsection{Simplification Methods}
\label{ssec:methods}
A core component of Alg.~\ref{alg:generalFlow} is the selection of
simplification strategy to apply
(Line~\ref{line:pickAttempts} in Alg.~\ref{alg:simplify}). 
We now describe our pool of neural network simplification methods, 
and the strategy that we suggest for
selecting among them. The goal of all the simplification methods we
propose here is to reduce neural network sizes, while keeping the network's
behavior (i.e., its outputs) similar to that of the original;
especially on the counter-example provided by either the faulty
verifier or the oracle verifier.
Note that a single simplification method can often be applied multiple times, in different ways, using different input parameters.

\mysubsection{Method 1: linearizing piecewise-linear activation
  functions between fully-connected layers.}  In general, the presence of activation functions is a
major source of complexity in the verification process of neural networks: they render the problem NP-complete, require complex mechanisms for linearly
approximating them, and often entail case-splitting that slows down
the verifiers~\cite{KaBaDiJuKo17,MuMaSiPuVe22,WaZhXuLiJaHsKo21}. Thus,
in order to simplify the neural network, we propose to eliminate such
activation functions, by \emph{fixing them to a single linear
  segment}, effectively replacing them with linear constraints.  This
procedure is performed on an entire layer at a time; which, in turn,
 creates a sequence of consecutive purely linear layers that can
then be merged into a single linear layer,  reducing the overall number of layers and neurons in  the network.

In choosing the linear segment to which each function is fixed, we
propose to use the counter-example $I$ provided by either the
faulty verifier or the oracle verifier. 
The output of the new linear segment we choose, with respect to $I$, will match the output 
of the activation function on $I$.  

For simplicity, we focus here
on the \relu{} activation function ($\relu(x)=\max{(x,0)}$), although
the technique is applicable to any piecewise-linear
function. Intuitively, in such cases we propose to replace \emph{active}
\relu{}s ($x\geq 0$) by the identify function, and \emph{inactive} \relu{}s ($x<0$)
 by zero. More formally, 
observe two consecutive layers, 
$l^t$ and $l^{t+1}$,
in the neural network $\mathcal{N}$,
where layer $l^t$ has a
\relu{} activation function. We construct an alternative layer,
$l^{a}$, to replace both $l^t$ and $l^{t+1}$. 
$l^{a}$ inherits the activation function of $l^{t+1}$. 
The weights 
\(W^{a}\) and 
the biases
\(B^{a}\) of $l^{a}$ are calculated as:
\begin{align*}
    W^{a}&= W^{t+1}W'W^{t}\\
    B^{a}&= W^{t+1}W'B^{t}+B^{t+1}
\end{align*}
where
\[W'_{i,j}=\begin{cases}
    1 & i=j \wedge \left(N^{l^t}_Q(I)\right)_{i}\geq 0\\
    0 & \text{otherwise}
\end{cases}
\]
Here $W'$ is the new linear segment replacing the activation function \relu{}. 
 Finally, the obtained simplified network
\(\mathcal{N}_{r}\) is the network \(\mathcal{N}\) where layers
\(l^t\) and \(l^{t+1}\) are deleted and replaced with \(l^{a}\).

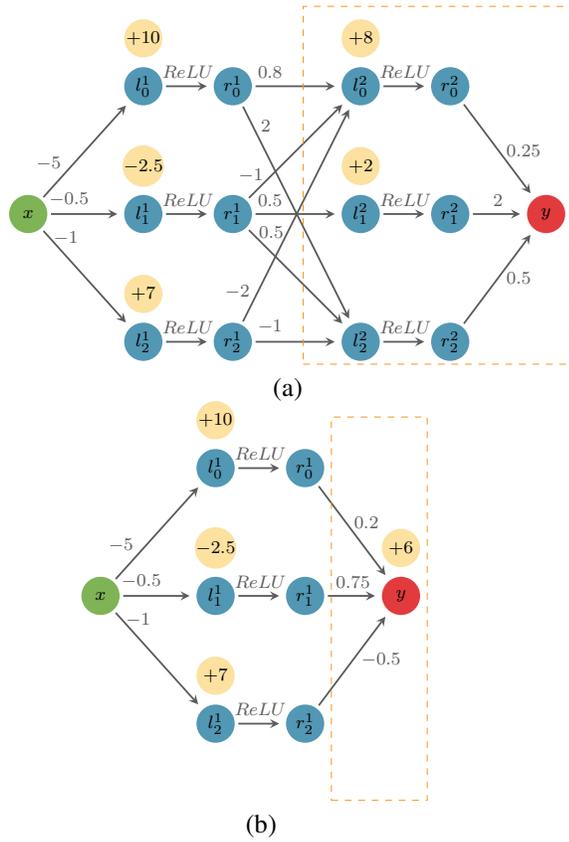
\begin{figure}[ht]
  \begin{minipage}{.4\textwidth}
    \scalebox{0.85}{
      \def\WSsep{2cm}
      \def\BSsep{-0.0cm}
      \def\IOWSsep{1.5cm}
      \def\relusep{1.4cm}
      \def\Ysep{-2}
      \begin{tikzpicture}[shorten >=1pt,->,draw=black!50, node distance=\layersep,font=\footnotesize]

        \node[input neuron] (I-0) at (0,1 * \Ysep) {$x$};

        \node[hidden neuron] (H-1-0) at (\IOWSsep + 0 * \relusep + 0 * \WSsep + 0.3cm, 0 * \Ysep) {$l^1_0$};
        \node[hidden neuron] (H-1-1) at (\IOWSsep + 0 * \relusep + 0 * \WSsep + 0.3cm, 1 * \Ysep) {$l^1_1$};
        \node[hidden neuron] (H-1-2) at (\IOWSsep + 0 * \relusep + 0 * \WSsep + 0.3cm, 2 * \Ysep) {$l^1_2$};

        \node[bias neuron] (B-1-0) at (\IOWSsep + 0 * \relusep + 0 * \WSsep + 0.3cm + \BSsep,0.75 + 0 * \Ysep) {$+10$};
        \node[bias neuron] (B-1-0) at (\IOWSsep + 0 * \relusep + 0 * \WSsep + 0.3cm + \BSsep,0.75 + 1 * \Ysep) {$-2.5$};
        \node[bias neuron] (B-1-0) at (\IOWSsep + 0 * \relusep + 0 * \WSsep + 0.3cm + \BSsep,0.75 + 2 * \Ysep) {$+7$};

        \node[hidden neuron] (R-1-0) at (\IOWSsep + 1 * \relusep + 0 * \WSsep + 0.3cm, 0 * \Ysep) {$r^1_0$};
        \node[hidden neuron] (R-1-1) at (\IOWSsep + 1 * \relusep + 0 * \WSsep + 0.3cm, 1 * \Ysep) {$r^1_1$};
        \node[hidden neuron] (R-1-2) at (\IOWSsep + 1 * \relusep + 0 * \WSsep + 0.3cm, 2 * \Ysep) {$r^1_2$};
        
        \node[hidden neuron] (H-2-0) at (\IOWSsep + 1 * \relusep + 1 * \WSsep + 0.3cm, 0 * \Ysep) {$l^2_0$};
        \node[hidden neuron] (H-2-1) at (\IOWSsep + 1 * \relusep + 1 * \WSsep + 0.3cm, 1 * \Ysep) {$l^2_1$};
        \node[hidden neuron] (H-2-2) at (\IOWSsep + 1 * \relusep + 1 * \WSsep + 0.3cm, 2 * \Ysep) {$l^2_2$};

        \node[bias neuron] (B-2-0) at (\IOWSsep + 1 * \relusep + 1 * \WSsep + 0.3cm + \BSsep,0.75) {$+8$};
        \node[bias neuron] (B-2-1) at (\IOWSsep + 1 * \relusep + 1 * \WSsep + 0.3cm + \BSsep,0.75 + 1 * \Ysep) {$+2$};

        \node[hidden neuron] (R-2-0) at (\IOWSsep + 2 * \relusep + 1 * \WSsep + 0.3cm, 0 * \Ysep) {$r^2_0$};
        \node[hidden neuron] (R-2-1) at (\IOWSsep + 2 * \relusep + 1 * \WSsep + 0.3cm, 1 * \Ysep) {$r^2_1$};
        \node[hidden neuron] (R-2-2) at (\IOWSsep + 2 * \relusep + 1 * \WSsep + 0.3cm, 2 * \Ysep) {$r^2_2$};
        
        \node[output neuron] (O-0) at (2 * \IOWSsep + 2 * \relusep + 1 * \WSsep + 0.3cm,1 * \Ysep) {$y$};

        \draw[nnedge] (I-0) --node[above,left,pos=0.37] {$-5$\;} (H-1-0);
        \draw[nnedge] (I-0) --node[above,pos=0.25] {\;$-0.5$} (H-1-1);
        \draw[nnedge] (I-0) --node[above,pos=0.25] {\;$-1$} (H-1-2);
        
        \draw[nnedge] (H-1-0) --node[above] {$\relu$} (R-1-0);
        \draw[nnedge] (H-1-1) --node[above] {$\relu$} (R-1-1);
        \draw[nnedge] (H-1-2) --node[above] {$\relu$} (R-1-2);
        
        \draw[nnedge] (R-1-0) --node[above,pos=0.2] {$0.8$} (H-2-0);
        \draw[nnedge] (R-1-0) --node[above,right,pos=0.1] {$2$} (H-2-2);
        
        \draw[nnedge] (R-1-1) --node[above,left,pos=0.25] {$-1$} (H-2-0);
        \draw[nnedge] (R-1-1) --node[above,pos=0.2] {$0.5$} (H-2-1);
        \draw[nnedge] (R-1-1) --node[above,right,pos=0.04] {$0.5$} (H-2-2);
        
        \draw[nnedge] (R-1-2) --node[above,left,pos=0.15] {$-2$} (H-2-0);
        \draw[nnedge] (R-1-2) --node[above,pos=0.2] {$-1$} (H-2-2);
        
        \draw[nnedge] (H-2-0) --node[above] {$\relu$} (R-2-0);
        \draw[nnedge] (H-2-1) --node[above] {$\relu$} (R-2-1);
        \draw[nnedge] (H-2-2) --node[above] {$\relu$} (R-2-2);
        
        \draw[nnedge] (R-2-0) --node[above,right] {$0.25$} (O-0);
        \draw[nnedge] (R-2-1) --node[above] {$2$} (O-0);
        \draw[nnedge] (R-2-2) --node[above,right] {$0.5$} (O-0);
        
        \draw[draw=color2,dashed] (\IOWSsep + 1 * \relusep + 0 * \WSsep + 0.3cm+1.1cm,1.25cm ) rectangle ++(2 * \IOWSsep + 2 * \relusep + 1 * \WSsep + 0.15cm -3.76cm, 3 * \Ysep + 0.4);

      \end{tikzpicture}
    }\centering\\\qquad\qquad(a)
  \end{minipage}
  \begin{minipage}{.4\textwidth}

      \centering
    \scalebox{0.85}{
      \def\WSsep{2cm}
      \def\BSsep{-0.0cm}
      \def\IOWSsep{1.5cm}
      \def\relusep{1.4cm}
      \def\Ysep{-2}
  \begin{tikzpicture}[shorten >=1pt,->,draw=black!50, node distance=\layersep,font=\footnotesize]

    \node[input neuron] (I-0) at (0,1 * \Ysep) {$x$};

    \node[hidden neuron] (H-1-0) at (\IOWSsep + 0 * \relusep + 0 * \WSsep + 0.3cm, 0 * \Ysep) {$l^1_0$};
    \node[hidden neuron] (H-1-1) at (\IOWSsep + 0 * \relusep + 0 * \WSsep + 0.3cm, 1 * \Ysep) {$l^1_1$};
    \node[hidden neuron] (H-1-2) at (\IOWSsep + 0 * \relusep + 0 * \WSsep + 0.3cm, 2 * \Ysep) {$l^1_2$};

    \node[bias neuron] (B-1-0) at (\IOWSsep + 0 * \relusep + 0 * \WSsep + 0.3cm + \BSsep,0.75 + 0 * \Ysep) {$+10$};
    \node[bias neuron] (B-1-0) at (\IOWSsep + 0 * \relusep + 0 * \WSsep + 0.3cm + \BSsep,0.75 + 1 * \Ysep) {$-2.5$};
    \node[bias neuron] (B-1-0) at (\IOWSsep + 0 * \relusep + 0 * \WSsep + 0.3cm + \BSsep,0.75 + 2 * \Ysep) {$+7$};

    \node[hidden neuron] (R-1-0) at (\IOWSsep + 1 * \relusep + 0 * \WSsep + 0.3cm, 0 * \Ysep) {$r^1_0$};
    \node[hidden neuron] (R-1-1) at (\IOWSsep + 1 * \relusep + 0 * \WSsep + 0.3cm, 1 * \Ysep) {$r^1_1$};
    \node[hidden neuron] (R-1-2) at (\IOWSsep + 1 * \relusep + 0 * \WSsep + 0.3cm, 2 * \Ysep) {$r^1_2$};

    \node[bias neuron] (B-O) at (2 * \IOWSsep + 1 * \relusep + 0 * \WSsep + 0.3cm + \BSsep,0.75 + 1 * \Ysep) {$+6$};

    \node[output neuron] (O-0) at (2 * \IOWSsep + 1 * \relusep + 0 * \WSsep + 0.3cm,1 * \Ysep) {$y$};
    
    \draw[nnedge] (I-0) --node[above,left,pos=0.37] {$-5$\;} (H-1-0);
    \draw[nnedge] (I-0) --node[above,pos=0.25] {\;$-0.5$} (H-1-1);
    \draw[nnedge] (I-0) --node[above,pos=0.25] {\;$-1$} (H-1-2);
    
    \draw[nnedge] (H-1-0) --node[above] {$\relu$} (R-1-0);
    \draw[nnedge] (H-1-1) --node[above] {$\relu$} (R-1-1);
    \draw[nnedge] (H-1-2) --node[above] {$\relu$} (R-1-2);
    
    \draw[nnedge] (R-1-0) --node[above,right,pos=0.4] {$0.2$} (O-0);    
    \draw[nnedge] (R-1-1) --node[above] {$0.75$} (O-0);
    \draw[nnedge] (R-1-2) --node[above,right] {$-0.5$} (O-0);
    
    \draw[draw=color2,dashed] (\IOWSsep + 1 * \relusep + 0 * \WSsep + 0.3cm+0.41cm,0.8cm) rectangle ++(2 * \IOWSsep + 1 * \relusep + 0 * \WSsep - 2.9cm,3 * \Ysep  );

  \end{tikzpicture}
  }\centering\\\qquad(b)

  \end{minipage}
  \caption{$\mathcal{N}_e$ with layers $l^2$ and $l^3$ selected in
    orange (a), and then merged (b).}
      \label{fig:reluLinearizationOne}
\end{figure}

Fig.~\ref{fig:reluLinearizationOne} depicts the result of 
applying this method on layers $l^2$ and $l^3$ from  
Fig.~\ref{fig:MainExample}, using 
the assignment $I_e = (5)$. 
Fig.~\ref{fig:reluLinearizationOne}a
 depicts the layers selected for merging; and Fig.~\ref{fig:reluLinearizationOne}b
depicts the resulting neural network. 
Notice that $\mathcal{N}_e^{l^2}(I_e) = (4, 2, -2)$, meaning that only
the \relu{}s in  neurons $l^2_0$ and $l^2_1$ are active. Thus, these
\relu{}s are replaced by the identity function, whereas the inactive
\relu{} of $l^2_2$ is replaced by $0$.
After this step, layers $l^2$ and $l^3$ perform only linear operations, and are
merged into a single layer. 

\medskip

\mysubsection{Method 2: linearizing piecewise-linear activation
  functions between convolutional layers.}
In this method, a convolutional layer is combined with the layer
following it (either a fully connected layer or a convolutional one),
and replaced by a single, fully connected layer. 


For simplicity, we focus here on the case where the second layer is
fully connected.  More formally, observe two consecutive layers, $l^t$
and $l^{t+1}$ in $\mathcal{N}$, where $l^t$ is a convolutional layer
and $l^{t+1}$ is a fully connected layer.  Our goal is to construct an
alternative layer, $l^{a}$, that will replace $l^t$ and $l^{t+1}$.
Since a convolutional layer is a particular case of fully connected
layer, we construct $l^a$ by first converting the convolutional layer
$l^t$ into a fully connected one, denoted $l^c$; then linearizing the activation
functions, as in \emph{Method 1}; and finally, combining the two
layers into one.



Denote by \(W^{t}\) and \(W^{{t+1}}\)  the matrices representing 
the weights of layer $l^t$ and $l^{t+1}$ respectively, 
and by \(B^{t}\) and \(B^{{t+1}}\)  the vectors representing 
their respective biases.
To transform  a convolutional layer into a fully connecting one, we calculate the weights, \(W^{c}\), and  the biases, \(B^{c}\),   of the fully connected layer replacing the convolutional one, according to the conventional layer parameters. First,  we  turn its input and output from a 
multidimensional tensors into 1-dimensional vectors.
The height and width (dimensions) of the feature maps in the convolutional layer's output are: $h_o ,  w_o $  where
\[
  h_o = \left\lfloor \frac{h + 2p - k}{ s}\right\rfloor + 1, \quad
  w_o = \left\lfloor \frac{w + 2p - k}{ s}\right\rfloor + 1 .
\]
The convolutional layer's output contains $c_o$ feature maps, i.e., the dimensions of the output are $(c_o\times h_o\times w_o)$. 
Thus, the dimensions of \(W^{c}\) are \(( c_o h_o w_o \times c_i h  w  )\).
\(W^{c}\) is a sparse matrix. To calculate the value of the \(i, j\)-th entry in  \(W^{c}\), 
 we first compute the following values: 
\begin{align*}
  c_i' &= \left\lfloor \frac{j}{hw}\right\rfloor, \quad
  c_o' = \left\lfloor \frac{i}{h_ow_o}\right\rfloor, \quad 
  \\
  i' &= \left\lfloor\frac{i - c_i  h  w}{w}\right\rfloor  - 
  \left(\left\lfloor\frac{j - c_o  h_o  w_o}{w_o}\right\rfloor  \cdot s - p\right)\\
  j' &=  ((i - c_i  h  w) \bmod{w})-\left(((j - c_o  h_o  w_o)\bmod{w_o}) \cdot s - p\right) \\
\end{align*}
$c_i'$ and $c_o'$ are the input and output channels that the \(i, j\)-th entry should be associated with.
$i'$ and $j'$ are the indices in the kernel that should match to the
\(i, j\)-th entry. The weight matrix $W^{c}$ is given by:
\[
W^{c}_{i,j} = \begin{cases}
     W^{t}_{c_i', c_o', i', j'} & 0 \leq i' \wedge j' < k \\ 
     W^{c}_{i,j} = 0 & \textit{otherwise}
\end{cases}
\]
Finally, \[B^c_i = B^{t}_{\left\lfloor\frac{i}{h_ow_o} \right\rfloor} \]

According to this construction of $W^c$ and $B^c$,  they will have the same functionality as the convolutional 
operation they replace. This step may temporarily increase the number of edges in 
the network (but not the number of neurons). This is required 
to prepare for the minimization step.

The next step is to linearize the \relu{}. 
This is done in a similar manner to the linearization in the previous method, from which we get $W'$.
Next, we construct the weights \(W^{a}\) and the biases \(B^{a}\) 
of the alternative layer $l^a$:
\begin{align*}
W^{a}=& W^{{t+1}}W'W^{c} \\
B^{a}=& W^{{t+1}}W'B^{c}+B^{{t+1}}
\end{align*}
And the activation function assigned to the new layer $l_{a}$ 
is the same as the one assigned to layer $l_{t+1}$. 
Finally, the simplified \nn{} \(\mathcal{N}_{r}\) is the network \(\mathcal{N}\),
where layers \(l_t\) and \(l_{t+1}\) are deleted and replaced with \(l_{a}\).

In case $l^{t+1}$ is also a convolutional layer, we convert it to a
fully connected layer, as we did with $l^t$; and the remainder of the
process is unchanged.

\mysubsection{Method 3: merging neurons.} In this method, we seek to
merge a pair of neurons in the same layer 
into a single neuron, thus decreasing the neural network
size by one. Of course, this entails selecting the weights of this new
neuron's incoming and outgoing edges, as well as its bias.
Our  motivation is to cause the merged neuron to produce values
close to those of the original neurons, and consequently cause
little changes in the neural network's eventual output. We present
first the technical process of merging neurons, and later discuss
\emph{which} pairs of neurons should be merged.

 We focus again on the case where the activation function is
\relu{}.  We first use the counter-example $I$ (returned by either the
faulty verifier or the oracle verifier) to check whether the
activation functions of the neurons being merged have the same phase
--- i.e., if they are both active, or both inactive. If they have the
same phase, we compute the merged neuron's weights and biases using the
original neurons' weights and biases. Specifically, the weight of each
edge incoming to the merged neuron is the mean of the original
incoming edge weights, and the neuron's bias is the mean of the original
neurons' biases; whereas the weights of its outgoing edges are the
weighted sum, according to $I$, of the original outgoing edge weights
(a weighted sum is needed, instead of a simple sum, to ensure that the
neurons in the following layer obtain values similar to their original
ones with respect to $I$). In case one of the neurons is active
and the other is inactive, we simply delete the inactive one,
since it does not contribute to the following layer's neuron values
(with respect to $I$).

Formally, given a neural network, $\mathcal{N}$, two successive layers in it, $l^t$ and $l^{t+1}$,
and two neurons indices $b<c$,  
we construct two alternative layers $l^{a}$ and $l^{a+1}$ that will replace $l^t$ and $l^{t+1}$ respectively. Additionally, 
$l^{a}$ and $l^{a+1}$ inherit the activation functions of $l^t$ and $l^{t+1}$ respectively.
If the \relu{}s of the neurons $b$ and $c$ in layer $l^t$ have the same phases: $\left(\mathcal{N}^{l^{t}}(I)\right)_b, \left(\mathcal{N}^{l^{t}}(I)\right)_c > 0$ or $\left(\mathcal{N}^{l^{t}}(I)\right)_b, \left(\mathcal{N}^{l^{t}}(I)\right)_c < 0$, the weights and the biases \(W^{{a}}, W^{{a+1}},  B^{{a}}, B^{{a+1}}\) of the alternative layers are calculated as follows:
\begin{align*}
B^{a}_{i}&=\begin{cases}
  B^{t}_{i} & i<b \vee b<i<c\\
  \frac{B^{t}_{b}+B^{t}_{c}}{2} & i = b \\
  B^{t}_{i+1} & c\leq i\\
\end{cases} \\
B^{{a+1}}&= B^{{t+1}} \\
W^{a}_{i,j}&=\begin{cases}
  W^{t}_{i,j} & i<b \vee b<i<c\\
  \frac{W^{t}_{b,j}+W^{t}_{c,j}}{2} & i = b \\
  W^{t}_{i+1,j} & c\leq i\\
\end{cases} 
\\
W^{{a+1}}_{i,j}&=\begin{cases}
    W^{{t+1}}_{i,j} \ \ \ \ \ \ \ \ \ \ \ \ \ \ \ \ \ \ \ \ \ \ \ \ \ \ \  j<b \vee b<j<c\\
    \frac{2
    \cdot \left(W^{{t+1}}_{i,b}
    \left(\mathcal{N}^{l^{t+1}}(I)\right)_{b}
    +W^{{t+1}}_{i,c}
    \left(\mathcal{N}^{l^{t+1}}(I)\right)_{c}
    \right)
    }{\left(\mathcal{N}^{l^{t+1}}(I)\right)_{b} + \left(\mathcal{N}^{l^{t+1}}(I)\right)_{c}} \ \ \  j = b \\
    W^{{t+1}}_{i,j+1} \ \ \ \ \ \ \ \ \ \ \ \ \ \ \ \ \ \ \ \ \ \ \ \ \ \ \ \ \  \ \ \ \ \ \ \ \ \ \ \ c\leq j\\
\end{cases} 
\end{align*}

Otherwise, if the \relu{}s of the neurons $b$ and $c$ in layer $l^t$ have different phases: $\left(\mathcal{N}^{l^{t}}(I)\right)_b > 0 \wedge \left(\mathcal{N}^{l^{t}}(I)\right)_c < 0$ (assume w.l.o.g. that the $c$-th neuron is the inactive one), 
the weights and biases \(W^{{a}}, W^{{a+1}},  B^{{a}}, B^{{a+1}}\) of the alternative layers are calculated as follows:
\begin{align*}
B^{a}_{i}=\begin{cases}
  B^{t}_{i} & i<c\\
  B^{t}_{i+1} & c\leq i\\
\end{cases}, \qquad
B^{{a+1}}= B^{{t+1}} \qquad  \qquad \quad\\
W^{a}_{i,j}=\begin{cases}
  W^{t}_{i,j} & i<c\\
  W^{t}_{i+1,j} & c\leq i\\
\end{cases}, \quad
W^{{a+1}}_{i,j}=\begin{cases}
    W^{{t+1}}_{i,j} & j<c\\
    W^{{t+1}}_{i,j+1} & c\leq j\\
\end{cases} 
\end{align*}

Finally, the obtained simplified neural network
\(\mathcal{N}_{r}\), is the network \(\mathcal{N}\)
where layers \(l^t\) and \(l^{t+1}\) are replaced with \(l^a
\) and \(l^{a+1}\) respectively.  This method can be applied
repeatedly, to reduce the network size even further.

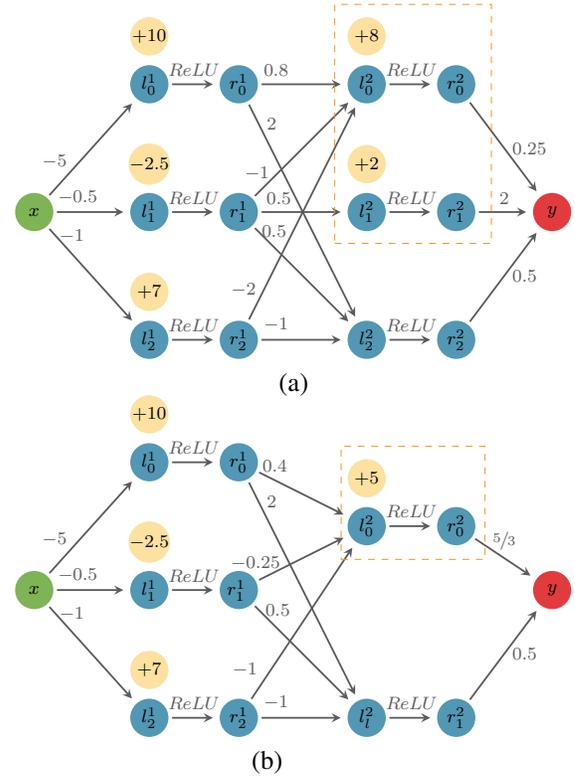
\begin{figure}[ht]
  \begin{minipage}{.4\textwidth}
    \scalebox{0.85}{
      \def\WSsep{2cm}
      \def\BSsep{-0.0cm}
      \def\IOWSsep{1.5cm}
      \def\relusep{1.4cm}
      \def\Ysep{-2}
      \begin{tikzpicture}[shorten >=1pt,->,draw=black!50, node distance=\layersep,font=\footnotesize]

        \node[input neuron] (I-0) at (0,1 * \Ysep) {$x$};

        \node[hidden neuron] (H-1-0) at (\IOWSsep + 0 * \relusep + 0 * \WSsep + 0.3cm,0 * \Ysep) {$l^1_0$};
        \node[hidden neuron] (H-1-1) at (\IOWSsep + 0 * \relusep + 0 * \WSsep + 0.3cm,1 * \Ysep) {$l^1_1$};
        \node[hidden neuron] (H-1-2) at (\IOWSsep + 0 * \relusep + 0 * \WSsep + 0.3cm,2 * \Ysep) {$l^1_2$};

        \node[bias neuron] (B-1-0) at (\IOWSsep + 0 * \relusep + 0 * \WSsep + 0.3cm + \BSsep,0.75 + 0 * \Ysep) {$+10$};
        \node[bias neuron] (B-1-0) at (\IOWSsep + 0 * \relusep + 0 * \WSsep + 0.3cm + \BSsep,0.75 + 1 * \Ysep) {$-2.5$};
        \node[bias neuron] (B-1-0) at (\IOWSsep + 0 * \relusep + 0 * \WSsep + 0.3cm + \BSsep,0.75 + 2 * \Ysep) {$+7$};

        \node[hidden neuron] (R-1-0) at (\IOWSsep + 1 * \relusep + 0 * \WSsep + 0.3cm, 0 * \Ysep) {$r^1_0$};
        \node[hidden neuron] (R-1-1) at (\IOWSsep + 1 * \relusep + 0 * \WSsep + 0.3cm, 1 * \Ysep) {$r^1_1$};
        \node[hidden neuron] (R-1-2) at (\IOWSsep + 1 * \relusep + 0 * \WSsep + 0.3cm, 2 * \Ysep) {$r^1_2$};
        
        \node[hidden neuron] (H-2-0) at (\IOWSsep + 1 * \relusep + 1 * \WSsep + 0.3cm, 0 * \Ysep) {$l^2_0$};
        \node[hidden neuron] (H-2-1) at (\IOWSsep + 1 * \relusep + 1 * \WSsep + 0.3cm, 1 * \Ysep) {$l^2_1$};
        \node[hidden neuron] (H-2-2) at (\IOWSsep + 1 * \relusep + 1 * \WSsep + 0.3cm, 2 * \Ysep) {$l^2_2$};

        \node[bias neuron] (B-2-0) at (\IOWSsep + 1 * \relusep + 1 * \WSsep + 0.3cm + \BSsep,0.75 + 0 * \Ysep) {$+8$};
        \node[bias neuron] (B-2-1) at (\IOWSsep + 1 * \relusep + 1 * \WSsep + 0.3cm + \BSsep,0.75 + 1 * \Ysep) {$+2$};

        \node[hidden neuron] (R-2-0) at (\IOWSsep + 2 * \relusep + 1 * \WSsep + 0.3cm, 0 * \Ysep) {$r^2_0$};
        \node[hidden neuron] (R-2-1) at (\IOWSsep + 2 * \relusep + 1 * \WSsep + 0.3cm, 1 * \Ysep) {$r^2_1$};
        \node[hidden neuron] (R-2-2) at (\IOWSsep + 2 * \relusep + 1 * \WSsep + 0.3cm, 2 * \Ysep) {$r^2_2$};
        
        \node[output neuron] (O-0) at (2 * \IOWSsep + 2 * \relusep + 1 * \WSsep + 0.3cm,1 * \Ysep) {$y$};
        
        \draw[nnedge] (I-0) --node[above,left,pos=0.37] {$-5$\;} (H-1-0);
        \draw[nnedge] (I-0) --node[above,pos=0.25] {\;\;$-0.5$} (H-1-1);
        \draw[nnedge] (I-0) --node[above,pos=0.25] {\;$-1$} (H-1-2);
        
        \draw[nnedge] (H-1-0) --node[above] {$\relu$} (R-1-0);
        \draw[nnedge] (H-1-1) --node[above] {$\relu$} (R-1-1);
        \draw[nnedge] (H-1-2) --node[above] {$\relu$} (R-1-2);
        
        \draw[nnedge] (R-1-0) --node[above,pos=0.2] {$0.8$} (H-2-0);
        \draw[nnedge] (R-1-0) --node[above,right,pos=0.1] {$2$} (H-2-2);
        
        \draw[nnedge] (R-1-1) --node[above,left,pos=0.25] {$-1$} (H-2-0);
        \draw[nnedge] (R-1-1) --node[above,pos=0.23] {$0.5$} (H-2-1);
        \draw[nnedge] (R-1-1) --node[above,pos=0.19] {\;$0.5$} (H-2-2);
        
        \draw[nnedge] (R-1-2) --node[above,left,pos=0.15] {$-2$} (H-2-0);
        \draw[nnedge] (R-1-2) --node[above,pos=0.2] {$-1$} (H-2-2);
        
        \draw[nnedge] (H-2-0) --node[above] {$\relu$} (R-2-0);
        \draw[nnedge] (H-2-1) --node[above] {$\relu$} (R-2-1);
        \draw[nnedge] (H-2-2) --node[above] {$\relu$} (R-2-2);
        
        \draw[nnedge] (R-2-0) --node[above,right] {$0.25$} (O-0);
        \draw[nnedge] (R-2-1) --node[above] {$2$} (O-0);
        \draw[nnedge] (R-2-2) --node[above,right] {$0.5$} (O-0);
        
        \draw[draw=color2,dashed] (\IOWSsep + 1 * \relusep + 0 * \WSsep + 0.3cm+1.5cm,1.25cm ) rectangle ++(1 * \IOWSsep + 2 * \relusep + 1 * \WSsep + 0.15cm -4cm,2 * \Ysep -3.6cm );

      \end{tikzpicture}
    }\centering\\\qquad\qquad(a)
  \end{minipage}
  \begin{minipage}{.4\textwidth}

      \centering
    \scalebox{0.85}{
      \def\WSsep{2cm}
      \def\BSsep{-0.0cm}
      \def\IOWSsep{1.5cm}
      \def\relusep{1.4cm}
      \def\Ysep{-2}
  \begin{tikzpicture}[shorten >=1pt,->,draw=black!50, node distance=\layersep,font=\footnotesize]

    \node[input neuron] (I-0) at (0,1 * \Ysep) {$x$};

    \node[hidden neuron] (H-1-0) at (\IOWSsep + 0 * \relusep + 0 * \WSsep + 0.3cm,0 * \Ysep) {$l^1_0$};
    \node[hidden neuron] (H-1-1) at (\IOWSsep + 0 * \relusep + 0 * \WSsep + 0.3cm,1 * \Ysep) {$l^1_1$};
    \node[hidden neuron] (H-1-2) at (\IOWSsep + 0 * \relusep + 0 * \WSsep + 0.3cm,2 * \Ysep) {$l^1_2$};

    \node[bias neuron] (B-1-0) at (\IOWSsep + 0 * \relusep + 0 * \WSsep + 0.3cm + \BSsep,0.75 + 0 * \Ysep) {$+10$};
    \node[bias neuron] (B-1-0) at (\IOWSsep + 0 * \relusep + 0 * \WSsep + 0.3cm + \BSsep,0.75 + 1 * \Ysep) {$-2.5$};
    \node[bias neuron] (B-1-0) at (\IOWSsep + 0 * \relusep + 0 * \WSsep + 0.3cm + \BSsep,0.75 + 2 * \Ysep) {$+7$};

    \node[hidden neuron] (R-1-0) at (\IOWSsep + 1 * \relusep + 0 * \WSsep + 0.3cm,0 * \Ysep) {$r^1_0$};
    \node[hidden neuron] (R-1-1) at (\IOWSsep + 1 * \relusep + 0 * \WSsep + 0.3cm,1 * \Ysep) {$r^1_1$\;};
    \node[hidden neuron] (R-1-2) at (\IOWSsep + 1 * \relusep + 0 * \WSsep + 0.3cm,2 * \Ysep) {$r^1_2$};
    
    \node[hidden neuron] (H-2-0) at (\IOWSsep + 1 * \relusep + 1 * \WSsep + 0.3cm, 0.5 * \Ysep) {$l^2_0$};
    \node[hidden neuron] (H-2-2) at (\IOWSsep + 1 * \relusep + 1 * \WSsep + 0.3cm, 2 * \Ysep) {$l^2_l$};

    \node[bias neuron] (B-2-0) at (\IOWSsep + 1 * \relusep + 1 * \WSsep + 0.3cm + \BSsep,0.75+ 0.5 * \Ysep) {$+5$};

    \node[hidden neuron] (R-2-0) at (\IOWSsep + 2 * \relusep + 1 * \WSsep + 0.3cm,0.5 * \Ysep) {$r^2_0$};
    \node[hidden neuron] (R-2-2) at (\IOWSsep + 2 * \relusep + 1 * \WSsep + 0.3cm,2 * \Ysep) {$r^2_1$};
    
    \node[output neuron] (O-0) at (2 * \IOWSsep + 2 * \relusep + 1 * \WSsep + 0.3cm,1 * \Ysep) {$y$};
    
    \draw[nnedge] (I-0) --node[above,left,pos=0.37] {$-5$\;} (H-1-0);
    \draw[nnedge] (I-0) --node[above,pos=0.25] {\;\;$-0.5$} (H-1-1);
    \draw[nnedge] (I-0) --node[above,pos=0.25] {\;$-1$} (H-1-2);
    
    \draw[nnedge] (H-1-0) --node[above] {$\relu$} (R-1-0);
    \draw[nnedge] (H-1-1) --node[above] {$\relu$} (R-1-1);
    \draw[nnedge] (H-1-2) --node[above] {$\relu$} (R-1-2);
    
    \draw[nnedge] (R-1-0) --node[above,pos=0.2] {$0.4$} (H-2-0);
    \draw[nnedge] (R-1-0) --node[above,right,pos=0.1] {$2$} (H-2-2);
    
    \draw[nnedge] (R-1-1) --node[above,left,pos=0.4] {$-0.25$\;} (H-2-0);
    \draw[nnedge] (R-1-1) --node[above,pos=0.2] {\;\;$0.5$} (H-2-2);
    
    \draw[nnedge] (R-1-2) --node[above,left,pos=0.2] {$-1\;$} (H-2-0);
    \draw[nnedge] (R-1-2) --node[above,pos=0.2] {$-1$} (H-2-2);
    
    \draw[nnedge] (H-2-0) --node[above] {$\relu$} (R-2-0);
    \draw[nnedge] (H-2-2) --node[above] {$\relu$} (R-2-2);
    
    \draw[nnedge] (R-2-0) --node[above] {$\sfrac{5}{3}$} (O-0);
    \draw[nnedge] (R-2-2) --node[above,right] {$0.5\;$} (O-0);
        
    \draw[draw=color2,dashed] (\IOWSsep + 1 * \relusep + 0 * \WSsep + 0.3cm+1.6cm,0.25cm ) rectangle ++(1 * \IOWSsep + 2 * \relusep + 1 * \WSsep + 0.15cm -4.2cm,1 * \Ysep-1.7cm );

  \end{tikzpicture}
  }\centering\\\qquad(b)
  \end{minipage}
  \caption{$\mathcal{N}_e$ with neurons $l^2_0$ and $l^2_1$ selected in
    orange (a), and then merged (b).}
  \label{fig:reluLinearization}
\end{figure}

An example of applying this method on the pair of neurons $l^2_0$ and $l^2_1$ in $\mathcal{N}_e$ 
from Fig.~\ref{fig:MainExample} using 
the assignment $I_e = (5)$  appears in Fig.~\ref{fig:reluLinearization}.
Fig.~\ref{fig:reluLinearization}a
shows the neurons selected for merging, and Fig.~\ref{fig:reluLinearization}b
shows the result of the merge.


\medskip

Choosing which pair of neurons to merge is crucial for the
success of this method. Every two neurons in the same layer are
valid candidates; 
however, some pairs are more likely to succeed than
others by resulting in a simplified neural network that behaves similarly to the original.
We consider the following possible approaches for prioritizing between
the pairs:
\begin{inparaenum}[(1)]
\item 
\label{app:arbitrarily} an arbitrary ordering;
\item
  \label{app:closest}
  prioritizing pairs with neurons that are assigned
  similar values (prior to the activation function), when the network
  is evaluated on assignment $I$.  The motivation is that
   merging such pairs is expected to have smaller effect on the overall
  functionality of the \nn{};
\item
  \label{app:negative} prioritizing pairs of neurons whose \relu{}s
  are inactive
  when  evaluated on $I$. The motivation is that inactive neurons may
  have
  little effect
  on the bug at hand. This approach can be combined with Approach~\ref{app:closest} to prioritize pairs with similar values after categorizing them by the status of the \relu{}s;
\item \label{app:posative} prioritizing pairs of neurons with positive values
  with respect to $I$. This approach, too, can be combined with
  Approach~\ref{app:closest}; and
\item \label{app:negativePosative}
  prioritizing pairs of neurons with negative values, and then pairs with positive values, with respect to $I$.
  This approach is a combination of
  Approaches 3 and 4, and again uses 
 Approach~\ref{app:closest} for internal prioritization within each category.
\end{inparaenum}

\mysubsection{Strategy for applying the simplification rules.} Within
Alg.~\ref{alg:generalFlow}, the simplification steps mentioned above
can be invoked in any order.  We propose to attempt methods that
significantly reduce the neural network size first, in order to reduce verification times.  We
empirically observed that this is achieved 
by the following strategy:
first, attempt to linearize and merge convolutional layers (\emph{Method 2}).
Second, attempt to linearize and merge fully connected layers
(\emph{Method 1}) --- starting with the output layer, and working backwards towards the input layer.
Finally, merge neurons  (\emph{Method 3}) according to
Approach~\ref{app:negativePosative}.
However, our implementation is highly customizable, and users can configure it
to use any other strategy, according to the task at hand.

To illustrate, applying our proposed strategy to $\mathcal{N}_e$ 
from Fig.~\ref{fig:MainExample},
with respect to the assignment $I_e = (5)$ in which
$\mathcal{N}_e^{l^1}(I_e) = (-15,-5, 2)$ and $\mathcal{N}_e^{l^2}(I_e) = (4, 2, -2)$, 
would result
in attempting the simplification methods in the following order:
\begin{enumerate*}[label=(\arabic*)]
  \item merge the layers $l^2$ and $l^3$;
  \item merge the layers $l^1$ and $l^2$;
  \item merge the pair of neurons \(l^1_0, l^1_1\);
  \item merge the pair of neurons \(l^2_1, l^2_2\); 
  \item merge the pair of neurons \(l^2_0, l^2_2\);
  \item merge the pair of neurons \(l^1_1, l^1_2\);  and then,
  \item merge the pair of neurons \(l^1_0, l^1_2\).
\end{enumerate*}
These steps are attempted, in order, until one succeeds; after which
the strategy is reapplied to the simplified network, and so on.

\section{Implementation and Evaluation}
\label{sec:evaluation}

We designed our tool, \ourTool{},  to be compatible with the standard input format used
in the VNN-COMP competition~\cite{https://doi.org/10.48550/arxiv.2109.00498}, in which
verification queries are encoded using the \emph{VNN-LIB} format~\cite{a0498}; and which,
in turn, relies on the \emph{Open Neural Network Exchange}
(\emph{ONNX}) format. This facilitated integrating \ourTool{}
with the various verifiers.  \ourTool{} is implemented in Python, and
contains classes that wrap objects of these formats. 
The tool has a modular design that allows applying
our proposed minimization methods in any order desired.


\sloppy VNN-COMP'21 included 12 participating \nnv{}s, and these were
tested on a set of verification queries.  We began by extracting from
the VNN-COMP'21 results pairs of dissenting verifiers, and the
verification queries that triggered these discrepancies. Each such
triple (two verifiers and a query) constitutes an input to
\ourTool{}. This extraction led us to target the following verifiers:
\begin{enumerate*}[label=(\arabic*)]
\item Marabou~\cite{katz2019marabou};
\item NNV~\cite{10.1007/978-3-030-53288-8_2, inbook23, tran2019star,
    https://doi.org/10.48550/arxiv.2004.05519,
    XiTrJo18};
\item
  NeuralVerification.jl (NV.jl)~\cite{LiArLaBaKo20}; and
\item nnenum~\cite{bak2020execution, 10.1007/978-3-030-53288-8_2, tran2019star, bak2021nnenum}.
\end{enumerate*} 
In the experiments described next, we used the same versions of these
verifiers that were used in VNN-COMP'21.

\mysubsection{Neuron Merging and Prioritization Approaches.}
For our first experiment, we set out to determine which of the
neuron-pair prioritization schemes described as part of \emph{Method 3} in
Sec.~\ref{ssec:methods} is the most successful. 
We measured success along two parameters: the size of the simplified
network obtained, and 
by the percentage of successful merging steps along the way.
We tested our algorithm on 5 input triples, involving networks of size 310 each.
Using only \emph{Method 3}, we ran \ourTool{}  with each of the
prioritization schemes, and counted for each, the number of
merging steps performed and the number of the steps that succeeded. 
Table.~\ref{reduction-merging-neurons} shows the results of this
comparison: the second column indicates, for every approach, the percentage
of the successful steps out of all the steps tried, aggregated
for all 5 benchmarks. 

Looking at the average reduction sizes, the results indicate that all
5 approaches were able to achieve a similar reduction in size, with a
slight advantage to approaches 1, 3 and 5.  However, the number of
successful merges varied significantly --- from Approach 1, in which
only 37.2\% of the merge steps were successful, and up to 75.9\%
for Approach 5 (in bold).  These results thus indicate that Approach 5 is the
most efficient of the 5, and so we used it as our default strategy for
Method 3 in the subsequent experiments.


\medskip
\begin{table}[ht]
\centering
\begin{center}
\caption{
  Comparing neurons merging approaches (Method 3) by
  size reduction and successful merges.}
\label{reduction-merging-neurons}
\begin{tabular}{ |l||c|c| } 
\hline 
 &  Successful merges (\%)& Average Reduction (\%)  \\
 \hline \hline 
 Approach 1 & 37.2\% & 96.0\% \\
 Approach 2 & 68.4\% & 95.9\% \\
 Approach 3 & 71.6\% & 96.0\% \\
 Approach 4 & 62.9\% & 95.8\% \\
 Approach 5 & \textbf{75.9\%} & 96.0\% \\
 \hline 
\end{tabular}
\end{center}
\end{table}

\mysubsection{Linearizing \relu{} Activations.}  In \emph{Method 1}
and \emph{Method 2} in Sec.~\ref{ssec:methods}, we proposed to
linearize activation functions, and then merge them with the
previous and following layers.  These methods can be applied to any
piecewise-linear activation function in the  network. The 
order in which they are applied is customizable.  In this experiment, we
set out to compare linearizing \relu{}s in ascending order (from input
layer towards output layer), and in descending order (from output
towards input).  Table~\ref{merging-layers-attempts} shows the results
of this experiment.

Every row in the table corresponds to an input triple to \ourTool{} (two disagreeing
verifiers and a verification query that they disagreed on), and the two
simplification approaches that were attempted. 
For each such experiment,
the second column indicates the number of simplification steps tried, until
\ourTool{} reached saturation (there were no additional steps to
try). The third column 
indicates the number of the successful steps out of all the 
steps. In column four,  the percentage  successful steps
out of all steps is shown;
and the final column shows the reduction percentage in the neural
network size. When one of the approaches was clearly superior, the
entry appears in bold.

To analyze the results, observe, e.g., the 5th experiment in
Table~\ref{merging-layers-attempts}. 
  The results imply that when using
the ascending approach, 12 linearizing and merging steps were made,
until the network count not be simplified further with either
\emph{Method 1} or \emph{Method 2}. Of these 12 steps, 5 were
successful --- and consequently, the simplified network has 5 fewer
layers than the original. In contrast, with the descending approach
only 9 steps were made until the network could not be simplified
further, 6 of which were successful.  Consequently, the simplified
network in this case has 6 fewer layers compared to the original.

The results  indicate that linearizing in descending order 
slightly outperforms linearizing
in ascending order, although the gap is not very significant. The
neural network in the last row included a convolutional layer, and,
according to the results, linearizing it in ascending order preformed
better.
After investigating this query further, we noticed that in the ascending 
order approach, the convolutional layer was merged into a fully connected 
one; whereas the descending approach did not succeed in removing 
or merging any convolutional layers.  We thus conclude that, for a
convolutional network, it is advisable to apply \emph{Method 2} before
applying \emph{Method 1}.

\begin{table}[ht]
\centering  
\caption{Comparing linearizing layers approaches by successful steps. * indicates the existence of a convolutional layer.\label{merging-layers-attempts}}
\begin{tabular}{ c |c|c|c|c|c|c|c| }
\cline{2-6}
&
                  			\begin{tabular}{@{}c@{}}Linearizing\\approach\end{tabular}         
                                & \begin{tabular}{@{}c@{}}No. of \\steps\end{tabular}
                                & \begin{tabular}{@{}c@{}c@{}}No. of\\successful\\steps\end{tabular}
                                & \begin{tabular}{@{}c@{}}Successful\\steps \%\end{tabular}
                                & \begin{tabular}{@{}c@{}}Neuron\\reduction \%\end{tabular}\\
\hline
\multirow{2}{*}{\begin{tabular}{@{}c@{}}1.\end{tabular}} 
& Ascending                     & 6             & 6             & 100.0\%       & 96.7\%        \\
& Descending                    & 6             & 6             & 100.0\%       & 96.7\%        \\
\hline
\multirow{2}{*}{\begin{tabular}{@{}c@{}}2.\end{tabular}} 
& Ascending                     & 6             & 6             & 100.0\%       & 96.7\%        \\
& Descending                    & 6             & 6             & 100.0\%       & 96.7\%        \\
\hline
\multirow{2}{*}{\begin{tabular}{@{}c@{}}3.\end{tabular}} 
& Ascending                     & 6             & 6             & 100.0\%       & 96.7\%        \\
& Descending                    & 6             & 6             & 100.0\%       & 96.7\%        \\
\hline
\multirow{2}{*}{\begin{tabular}{@{}c@{}}4.\end{tabular}} 
& Ascending                     & 6             & 0             & 0.0\%         & 0.0\%         \\
& Descending                    & 6             & 0             & 0.0\%         & 0.0\%         \\
\hline
\multirow{2}{*}{\begin{tabular}{@{}c@{}}5.\end{tabular}} 
& Ascending                     & 12            & 5             & 41.6\%        & 80.6\%        \\
& Descending                    & 9             & 6             & \bf{66.6}\%   & \bf{96.7}\%   \\
\hline
\multirow{2}{*}{\begin{tabular}{@{}c@{}}6.\end{tabular}} 
& Ascending                     & 3             & 2             & 66.6\%        & 39.2\%        \\
& Descending                    & 2             & 2             & \bf{100.0}\%  & 39.2\%        \\
\hline 
\multirow{2}{*}{\begin{tabular}{@{}c@{}}7.\end{tabular}} 
& Ascending                     & 3*             & 2*             & \bf{66.6}\%   & \bf{65.8}\%   \\ 
& Descending                    & 2*             & 1             & 50.0\%        & 0.0\%         \\
\hline
\end{tabular}
\end{table}

\mysubsection{Delta Debugging Discrepancies from VNN-COMP'21.}  For
our final experiment, we considered 13 triples of verifiers, oracle
verifiers, and verification queries. Of these triples, 11 contained
DNNs from the ACAS-Xu family~\cite{KaBaDiJuKo17}, 1 was a DNN from the
MNIST DNNs~\cite{Le98}, and 1 was a DNN from the Oval21
benchmark~\cite{https://doi.org/10.48550/arxiv.2109.00498}.  Using the
optimal configuration of our tool as previously discussed, we applied
the full-blown delta-debugging algorithm to all of our 13
benchmarks. The results appear in Table.~\ref{final-results}.  Every
row in the table represents a triple, and the first two columns
indicate the number of neurons in the original network, and the number
of remaining neurons after delta debugging was applied. The next two
columns indicate the number of layers in the original and reduced
networks; and the final column indicates the percent of neurons that
were removed.

\begin{table}[h]
\centering
\caption{Delta-debugging using our algorithm.
    * indicates the existence of a convolutional layer.}
\label{final-results}
\begin{tabular}{ |c|c||c|c||c|c||c| } 
  \hline
\multicolumn{2}{|c||}{\ Neurons\ }             &\multicolumn{2}{|c||}{Layers}        &\multirow{2}{*}{\begin{tabular}{@{}c@{}}Reduction \\ percentage\end{tabular}}    \\
\cline{1-4}
     In Original        & In reduced        & In original          & In reduced               &                                                                                 \\      
\hline\hline  
310                   & 6                     & 8                     & 2                           & 98\%                                                                            \\ 
310                   & 7                     & 8                     & 2                           & 97\%                                                                            \\ 
310                   & 6                     & 8                     & 2                           & 98\%                                                                            \\ 
310                   & 12                    & 8                     & 8                           & 96\%                                                                            \\ 
310                   & 6                     & 8                     & 2                           & 98\%                                                                            \\ 
9326                  & 12                    & 5*                    & 3                           & 99\%                                                                            \\ 
1306                  & 11                    & 4                     & 2                           & 99\%                                                                            \\ 
310                   & 10                    & 8                     & 3                           & 96\%                                                                            \\ 
310                   & 6                     & 8                     & 2                           & 98\%                                                                            \\ 
310                   & 10                    & 8                     & 4                           & 96\%                                                                            \\ 
310                   & 10                    & 8                     & 4                           & 96\%                                                                            \\ 
310                   & 9                     & 8                     & 4                           & 97\%                                                                            \\ 
310                   & 13                    & 8                     & 6                           & 95\%                                                                            \\ 
  \hline
\end{tabular}
\end{table}

Overall, the algorithm performed exceedingly well, reducing the
network sizes by an average of 96.8\% (!); and, in some cases, causing a
size decrease of 99\%, from a neural network with 1306 neurons and 4 layers
to just 11 neurons and 2 layers (an input layer and an output layer, without any activation functions). 
The minimal decrease observed was
95\%, from 310 neurons to 13. We regard these results as a very strong
indication of the usefulness of delta debugging in the context of DNN
verification.  Further analyzing the results, we observe that the
\relu{} linearization simplification rule was responsible for an
average of 66\% of the size reduction, whereas the remaining two rules
were responsible for an average of 34\% --- indicating that 
the \relu{} linearization simplification rule is the
main workhorse of our approach at its current configuration.

\section{Related Work}
\label{sec:relatedWork}
With the increasing pervasiveness of DNNs, the verification community
has been devoting growing efforts to verifying them. Numerous
approaches have been proposed, including SMT-based
approaches~\cite{KaBaDiJuKo17, KaBaDiJuKo21,
  katz2019marabou, GoKaPaBa18, WuOzZeIrJuGoFoKaPaBa20, StWuZeJuKaBaKo21},
approaches based on LP or MILP solvers~\cite{Eh17, TjXiTe17,
  BuTuToKoMu18}, reachability-based approaches~\cite{LoMa17,
  XiTrJo18}, abstraction and abstract-interpretation based
approaches~\cite{AsHaKrMo20, PrAf20, huang2017safety, MuMaSiPuVe22,
  SiGePuVe19, GeMiDrTsChVe18, GoPaPuRuSa21, WaZhXuLiJaHsKo21},
synthesis-based approaches~\cite{KoLoJaBl20, PoAbKr20}, run-time
optimization~\cite{AvBlChHeKoPr19, AnPaDiCh19}, quantitative
verification~\cite{BaShShMeSa19}, verification of recurrent
networks~\cite{ZhShGuGuLeNa20, JaBaKa20}, and many others. These
approaches, in turn, have been used in numerous application
domains~\cite{UrChWuZh20, SoTh19, YaYaTrHoJoPo21, GoAdKeKa20,
  DoSuWaWaDa20, UsGoSuNoPa21, ElKaKaSc21}. Given the scope of these efforts, and
the number of available tools, it is not surprising that bugs are
abundant, and that engineers are in need of efficient debugging tools.

To the best of our knowledge, no previous work has applied delta
debugging in the context of DNN verification, although similar
approaches have been shown successful in the related domains of
SMT~\cite{10.1145/1670412.1670413, NiPrBa22} and SAT~\cite{BrLoBi10}
solving. Related efforts have attempted to reduce DNN sizes, with the
purpose of producing smaller-but-equivalent networks, or networks
smaller with a respect to a particular verification property of
interest~\cite{PrAf20, Pr22, AsHaKrMo20, LaKa21}. In the future,
principles from these approaches could be integrated as simplification
strategies within our delta-debugging approach.

\section{Conclusion}
\label{sec:conclusion}

In this paper, we presented the \ourTool{} tool for automatically reducing the
size of a \vq{} with respect to an erroneous \nnv{}.  We focused on
delta-debugging techniques, and proposed multiple minimization methods for
reducing \nn{} sizes. These techniques attempt to simplify the neural
network in question, while modifying it as little as possible.  
We also suggested a strategy for the order in which to apply those methods.
We demonstrated
the effectiveness of \ourTool{} on actual benchmarks from the VNN-COMP'21
competition, and were able to significantly simplify them.
We regard this work as another step towards more sound tools for 
DNN verification.

\mysubsection{Acknowledgements.}  This work was 
	partially supported by the Israel Science Foundation (grant number
	683/18).

\newpage

\bibliographystyle{abbrv}
\bibliography{ref}

\begin{thebibliography}{10}

\bibitem{Al21}
A.~Albarghouthi.
\newblock {\em {Introduction to Neural Network Verification}}.
\newblock verifieddeeplearning.com, 2021.

\bibitem{AmFrKaMaRe23}
G.~Amir, Z.~Freund, G.~Katz, E.~Mandelbaum, and I.~Refaeli.
\newblock {veriFIRE: Verifying an Industrial, Learning-Based Wildfire
  Detection}.
\newblock In {\em Proc. 25th Int. Symposium on Formal Methods (FM)}, pages
  648--656, 2023.

\bibitem{AmKaSc22}
G.~Amir, G.~Katz, and M.~Schapira.
\newblock {Verification-Aided Deep Ensemble Selection}.
\newblock In {\em Proc. 22nd Int. Conf. on Formal Methods in Computer-Aided
  Design (FMCAD)}, pages 27--37, 2022.

\bibitem{AmScKa21}
G.~Amir, M.~Schapira, and G.~Katz.
\newblock {Towards Scalable Verification of Deep Reinforcement Learning}.
\newblock In {\em Proc. 21st Int. Conf. on Formal Methods in Computer-Aided
  Design (FMCAD)}, pages 193--203, 2021.

\bibitem{AnPaDiCh19}
G.~Anderson, S.~Pailoor, I.~Dillig, and S.~Chaudhuri.
\newblock {Optimization and Abstraction: a Synergistic Approach for Analyzing
  Neural Network Robustness}.
\newblock In {\em Proc. 40th ACM SIGPLAN Conf. on Programming Languages Design
  and Implementations (PLDI)}, pages 731--744, 2019.

\bibitem{AsHaKrMo20}
P.~Ashok, V.~Hashemi, J.~Kretinsky, and S.~Mohr.
\newblock {DeepAbstract: Neural Network Abstraction for Accelerating
  Verification}.
\newblock In {\em Proc. 18th Int. Symp. on Automated Technology for
  Verification and Analysis (ATVA)}, pages 92--107, 2020.

\bibitem{AvBlChHeKoPr19}
G.~Avni, R.~Bloem, K.~Chatterjee, T.~Henzinger, B.~K\"{o}nighofer, and
  S.~Pranger.
\newblock {Run-Time Optimization for Learned Controllers through Quantitative
  Games}.
\newblock In {\em Proc. 31st Int. Conf. on Computer Aided Verification (CAV)},
  pages 630--649, 2019.

\bibitem{bak2021nnenum}
B.~Bak.
\newblock {nnenum: Verification of Relu Neural Networks with Optimized
  Abstraction Refinement}.
\newblock In {\em Proc. 13th NASA Formal Methods Symposium (NFM)}, pages
  19--36, 2021.

\bibitem{bak2020execution}
S.~Bak.
\newblock {Execution-Guided Overapproximation (EGO) for Improving Scalability
  of Neural Network Verification}.
\newblock In {\em Proc. 3rd Int. Workshop on Verification of Neural Networks
  (VNN)}, 2020.

\bibitem{https://doi.org/10.48550/arxiv.2109.00498}
S.~Bak, C.~Liu, and T.~Johnson.
\newblock {The Second International Verification of Neural Networks Competition
  (VNN-COMP 2021): Summary and Results}, 2021.
\newblock Technical Report. \url{http://arxiv.org/abs/2109.00498}.

\bibitem{BaShShMeSa19}
T.~Baluta, S.~Shen, S.~Shinde, K.~Meel, and P.~Saxena.
\newblock {Quantitative Verification of Neural Networks and its Security
  Applications}.
\newblock In {\em Proc. ACM SIGSAC Conf. on Computer and Communications
  Security (CCS)}, pages 1249--1264, 2019.

\bibitem{a0498}
C.~Barrett, G.~Katz, D.~Guidotti, L.~Pulina, N.~Narodytska, and A.~Tacchella.
\newblock {The Verification of Neural Networks Library (VNN-LIB)}, 2019.
\newblock \url{www.vnnlib.org}.

\bibitem{10.1145/1670412.1670413}
R.~Brummayer and A.~Biere.
\newblock {Fuzzing and Delta-Debugging SMT Solvers}.
\newblock In {\em Proc. 7th Int. Workshop on Satisfiability Modulo Theories
  (SMT)}, 2009.

\bibitem{BrLoBi10}
R.~Brummayer, F.~Lonsing, and A.~Biere.
\newblock {Automated Testing and Debugging of SAT and QBF Solvers}.
\newblock In {\em Proc. 13th Int. Conf. on Theory and Applications of
  Satisfiability Testing (SAT)}, pages 44--57, 2010.

\bibitem{BuTuToKoMu18}
R.~Bunel, I.~Turkaslan, P.~Torr, P.~Kohli, and P.~Mudigonda.
\newblock {A Unified View of Piecewise Linear Neural Network Verification}.
\newblock In {\em Proc. 32nd Conf. on Neural Information Processing Systems
  (NeurIPS)}, pages 4795--4804, 2018.

\bibitem{DoSuWaWaDa20}
G.~Dong, J.~Sun, J.~Wang, X.~Wang, and T.~Dai.
\newblock {Towards Repairing Neural Networks Correctly}, 2020.
\newblock Technical Report. \url{http://arxiv.org/abs/2012.01872}.

\bibitem{Eh17}
R.~Ehlers.
\newblock {Formal Verification of Piece-Wise Linear Feed-Forward Neural
  Networks}.
\newblock In {\em Proc. 15th Int. Symp. on Automated Technology for
  Verification and Analysis (ATVA)}, pages 269--286, 2017.

\bibitem{ElKaKaSc21}
T.~Eliyahu, Y.~Kazak, G.~Katz, and M.~Schapira.
\newblock {Verifying Learning-Augmented Systems}.
\newblock In {\em Proc. Conf. of the ACM Special Interest Group on Data
  Communication on the Applications, Technologies, Architectures, and Protocols
  for Computer Communication (SIGCOMM)}, pages 305--318, 2021.

\bibitem{GeMiDrTsChVe18}
T.~Gehr, M.~Mirman, D.~Drachsler-Cohen, E.~Tsankov, S.~Chaudhuri, and
  M.~Vechev.
\newblock {AI2: Safety and Robustness Certification of Neural Networks with
  Abstract Interpretation}.
\newblock In {\em Proc. 39th IEEE Symposium on Security and Privacy (S\&P)},
  2018.

\bibitem{goldberg2016primer}
Y.~Goldberg.
\newblock {A Primer on Neural Network Models for Natural Language Processing}.
\newblock {\em Journal of Artificial Intelligence Research}, 57:345--420, 2016.

\bibitem{GoAdKeKa20}
B.~Goldberger, Y.~Adi, J.~Keshet, and G.~Katz.
\newblock {Minimal Modifications of Deep Neural Networks using Verification}.
\newblock In {\em Proc. 23rd Int. Conf. on Logic for Programming, Artificial
  Intelligence and Reasoning (LPAR)}, pages 260--278, 2020.

\bibitem{Goodfellow-et-al-2016}
I.~Goodfellow, Y.~Bengio, and A.~Courville.
\newblock {\em Deep Learning}.
\newblock MIT Press, 2016.

\bibitem{43405}
I.~Goodfellow, J.~Shlens, and C.~Szegedy.
\newblock {Explaining and Harnessing Adversarial Examples}, 2014.
\newblock Technical Report. \url{http://arxiv.org/abs/1412.6572}.

\bibitem{GoKaPaBa18}
D.~Gopinath, G.~Katz, C.~P\v{a}s\v{a}reanu, and C.~Barrett.
\newblock {DeepSafe: A Data-driven Approach for Checking Adversarial Robustness
  in Neural Networks}.
\newblock In {\em Proc. 16th. Int. Symp. on on Automated Technology for
  Verification and Analysis (ATVA)}, pages 3--19, 2018.

\bibitem{GoPaPuRuSa21}
E.~Goubault, S.~Palumby, S.~Putot, L.~Rustenholz, and S.~Sankaranarayanan.
\newblock {Static Analysis of ReLU Neural Networks with Tropical Polyhedra}.
\newblock In {\em Proc. 28th Int. Symposium on Static Analysis (SAS)}, pages
  166--190, 2021.

\bibitem{guo2020gluoncv}
J.~Guo, H.~He, T.~He, L.~Lausen, M.~Li, H.~Lin, X.~Shi, C.~Wang, J.~Xie,
  S.~Zha, et~al.
\newblock {GluonCV and GluonNLP: Deep Learning in Computer Vision and Natural
  Language Processing}.
\newblock {\em Journal of Machine Learning Research}, 21(23):1--7, 2020.

\bibitem{hou2018deepsf}
J.~Hou, B.~Adhikari, and J.~Cheng.
\newblock {DeepSF: Deep Convolutional Neural Network for Mapping Protein
  Sequences to Folds}.
\newblock {\em Bioinformatics}, 34(8):1295--1303, 2018.

\bibitem{huang2017safety}
X.~Huang, M.~Kwiatkowska, S.~Wang, and M.~Wu.
\newblock {Safety Verification of Deep Neural Networks}.
\newblock In {\em Proc. 29th Int. Conf. on Computer Aided Verification (CAV)},
  pages 3--29, 2017.

\bibitem{JaBaKa20}
Y.~Jacoby, C.~Barrett, and G.~Katz.
\newblock {Verifying Recurrent Neural Networks using Invariant Inference}.
\newblock In {\em Proc. 18th Int. Symposium on Automated Technology for
  Verification and Analysis (ATVA)}, pages 57--74, 2020.

\bibitem{JiRi20b}
K.~Jia and M.~Rinard.
\newblock {Exploiting Verified Neural Networks via Floating Point Numerical
  Error}, 2020.
\newblock Technical Report. \url{http://arxiv.org/abs/2003.03021}.

\bibitem{KaBaDiJuKo17}
G.~Katz, C.~Barrett, D.~Dill, K.~Julian, and M.~Kochenderfer.
\newblock {Reluplex: An Efficient SMT Solver for Verifying Deep Neural
  Networks}.
\newblock In {\em Proc. 29th Int. Conf. on Computer Aided Verification (CAV)},
  pages 97--117, 2017.

\bibitem{KaBaDiJuKo21}
G.~Katz, C.~Barrett, D.~Dill, K.~Julian, and M.~Kochenderfer.
\newblock {Reluplex: a Calculus for Reasoning about Deep Neural Networks},
  2021.

\bibitem{katz2019marabou}
G.~Katz, D.~Huang, D.~Ibeling, K.~Julian, C.~Lazarus, R.~Lim, P.~Shah,
  S.~Thakoor, H.~Wu, A.~Zelji\'c, D.~Dill, M.~Kochenderfer, and C.~Barrett.
\newblock {The Marabou Framework for Verification and Analysis of Deep Neural
  Networks}.
\newblock In {\em Proc. 31st Int. Conf. on Computer Aided Verification (CAV)},
  pages 443--452, 2019.

\bibitem{KoLoJaBl20}
B.~K\"{o}nighofer, F.~Lorber, N.~Jansen, and R.~Bloem.
\newblock {Shield Synthesis for Reinforcement Learning}.
\newblock In {\em Proc. Int. Symposium on Leveraging Applications of Formal
  Methods, Verification and Validation (ISoLA)}, pages 290--306, 2020.

\bibitem{LaKa21}
O.~Lahav and G.~Katz.
\newblock {Pruning and Slicing Neural Networks using Formal Verification}.
\newblock In {\em Proc. 21st Int. Conf. on Formal Methods in Computer-Aided
  Design (FMCAD)}, pages 183--192, 2021.

\bibitem{Le98}
Y.~LeCun.
\newblock {The MNIST Database of Handwritten Digits}, 1998.
\newblock \url{http://yann.lecun.com/exdb/mnist/}.

\bibitem{LiArLaBaKo20}
C.~Liu, T.~Arnon, C.~Lazarus, C.~Barrett, and M.~Kochenderfer.
\newblock {Algorithms for Verifying Deep Neural Networks}, 2020.
\newblock Technical Report. \url{http://arxiv.org/abs/1903.06758}.

\bibitem{liu2019multi}
X.~Liu, P.~He, W.~Chen, and J.~Gao.
\newblock {Multi-Task Deep Neural Networks for Natural Language Understanding},
  2019.
\newblock Technical Report. \url{http://arxiv.org/abs/1901.11504}.

\bibitem{LoMa17}
A.~Lomuscio and L.~Maganti.
\newblock {An Approach to Reachability Analysis for Feed-Forward ReLU Neural
  Networks}, 2017.
\newblock Technical Report. \url{http://arxiv.org/abs/1706.07351}.

\bibitem{MuMaSiPuVe22}
M.~M\"uller, G.~Makarchuk, G.~Singh, M.~P\"uschel, and M.~Vechev.
\newblock {PRIMA: General and Precise Neural Network Certification via Scalable
  Convex Hull Approximations}.
\newblock In {\em Proc. 49th ACM SIGPLAN Symposium on Principles of Programming
  Languages (POPL)}, 2022.

\bibitem{NiPrBa22}
A.~Niemetz, M.~Preiner, and C.~Barrett.
\newblock {Murxla: A Modular and Highly Extensible API Fuzzer for SMT Solvers}.
\newblock In {\em Proc. 34th Int. Conf. on Computer Aided Verification (CAV)},
  pages 92--106, 2022.

\bibitem{noe2020machine}
F.~No{\'e}, G.~De~Fabritiis, and C.~Clementi.
\newblock {Machine Learning for Protein Folding and Dynamics}.
\newblock {\em Current Opinion in Structural Biology}, 60:77--84, 2020.

\bibitem{OsBaKa22}
M.~Ostrovsky, C.~Barrett, and G.~Katz.
\newblock {An Abstraction-Refinement Approach to Verifying Convolutional Neural
  Networks}.
\newblock In {\em Proc. 20th. Int. Symposium on Automated Technology for
  Verification and Analysis (ATVA)}, pages 391--396, 2022.

\bibitem{PoAbKr20}
E.~Polgreen, R.~Abboud, and D.~Kroening.
\newblock {Counterexample Guided Neural Synthesis}, 2020.
\newblock Technical Report. \url{https://arxiv.org/abs/2001.09245}.

\bibitem{Pr22}
P.~Prabhakar.
\newblock {Bisimulations for Neural Network Reduction}.
\newblock In {\em Proc. 23rd Int. Conf. Verification on Model Checking, and
  Abstract Interpretation (VMCAI)}, pages 285--300, 2022.

\bibitem{PrAf20}
P.~Prabhakar and Z.~Afzal.
\newblock {Abstraction Based Output Range Analysis for Neural Networks}, 2020.
\newblock Technical Report. \url{https://arxiv.org/abs/2007.09527}.

\bibitem{PuTa10}
L.~Pulina and A.~Tacchella.
\newblock {An Abstraction-Refinement Approach to Verification of Artificial
  Neural Networks}.
\newblock In {\em Proc. 22nd Int. Conf. on Computer Aided Verification (CAV)},
  pages 243--257, 2010.

\bibitem{SiGePuVe19}
G.~Singh, T.~Gehr, M.~Puschel, and M.~Vechev.
\newblock {An Abstract Domain for Certifying Neural Networks}.
\newblock In {\em Proc. 46th ACM SIGPLAN Symposium on Principles of Programming
  Languages (POPL)}, 2019.

\bibitem{SoTh19}
M.~Sotoudeh and A.~Thakur.
\newblock {Correcting Deep Neural Networks with Small, Generalizing Patches}.
\newblock In {\em Workshop on Safety and Robustness in Decision Making}, 2019.

\bibitem{StWuZeJuKaBaKo21}
C.~Strong, H.~Wu, A.~Zelji\'c, K.~Julian, G.~Katz, C.~Barrett, and
  M.~Kochenderfer.
\newblock {Global Optimization of Objective Functions Represented by ReLU
  Networks}.
\newblock {\em Journal of Machine Learning}, pages 1--28, 2021.

\bibitem{TjXiTe17}
V.~Tjeng, K.~Xiao, and R.~Tedrake.
\newblock {Evaluating Robustness of Neural Networks with Mixed Integer
  Programming}, 2017.
\newblock Technical Report. \url{http://arxiv.org/abs/1711.07356}.

\bibitem{10.1007/978-3-030-53288-8_2}
H.-D. Tran, S.~Bak, W.~Xiang, and T.~Johnson.
\newblock {Verification of Deep Convolutional Neural Networks Using
  ImageStars}.
\newblock In {\em Proc. 32nd Int. Conf. on Computer Aided Verification (CAV)},
  pages 18--42, 2020.

\bibitem{tran2019star}
H.-D. Tran, D.~Manzanas~Lopez, P.~Musau, X.~Yang, L.~Nguyen, W.~Xiang, and
  T.~Johnson.
\newblock {Star-Based Reachability Analysis of Deep Neural Networks}.
\newblock In {\em Proc. Int. Symposium on Formal Methods (FM)}, pages 670--686,
  2019.

\bibitem{inbook23}
H.-D. Tran, P.~Musau, D.~Lopez, X.~Yang, L.~Nguyen, W.~Xiang, and T.~Johnson.
\newblock {Parallelizable Reachability Analysis Algorithms for Feed-Forward
  Neural Networks}.
\newblock In {\em Proc. 7th Int. Workshop on Formal Methods in Software
  Engineering (FormaliSE)}, pages 31--40, 2019.

\bibitem{https://doi.org/10.48550/arxiv.2004.05519}
H.-D. Tran, X.~Yang, D.~Lopez, P.~Musau, L.~Nguyen, W.~Xiang, S.~Bak, and
  T.~Johnson.
\newblock {NNV: The Neural Network Verification Tool for Deep Neural Networks
  and Learning-Enabled Cyber-Physical Systems}, 2020.
\newblock Technical Report. \url{http://arxiv.org/abs/2004.05519}.

\bibitem{UrChWuZh20}
C.~Urban, M.~Christakis, V.~W{\"u}stholz, and F.~Zhang.
\newblock {Perfectly Parallel Fairness Certification of Neural Networks}.
\newblock In {\em Proc. ACM Int. Conf. on Object Oriented Programming Systems
  Languages and Applications (OOPSLA)}, pages 1--30, 2020.

\bibitem{UsGoSuNoPa21}
M.~Usman, D.~Gopinath, Y.~Sun, Y.~Noller, and C.~P\v{a}s\v{a}reanu.
\newblock {NNrepair: Constraint-based Repair of Neural Network Classifiers},
  2021.
\newblock Technical Report. \url{http://arxiv.org/abs/2103.12535}.

\bibitem{nnvcomp20}
{International Verification of Neural Networks Competition (VNN-COMP)}, 2020.
\newblock \url{https://sites.google.com/view/vnn20/vnncomp}.

\bibitem{WaZhXuLiJaHsKo21}
S.~Wang, H.~Zhang, K.~Xu, X.~Lin, S.~Jana, C.-J. Hsieh, and Z.~Kolter.
\newblock {Beta-CROWN: Efficient Bound Propagation with Per-Neuron Split
  Constraints for Complete and Incomplete Neural Network Verification}.
\newblock In {\em Proc. 35th Conf. on Neural Information Processing Systems
  (NeurIPS)}, 2021.

\bibitem{WuOzZeIrJuGoFoKaPaBa20}
H.~Wu, A.~Ozdemir, A.~Zelji\'c, A.~Irfan, K.~Julian, D.~Gopinath, S.~Fouladi,
  G.~Katz, C.~P\u{a}s\u{a}reanu, and C.~Barrett.
\newblock {Parallelization Techniques for Verifying Neural Networks}.
\newblock In {\em Proc. 20th Int. Conf. on Formal Methods in Computer-Aided
  Design (FMCAD)}, pages 128--137, 2020.

\bibitem{WuZeKaBa22}
H.~Wu, A.~Zelji\'c, G.~Katz, and C.~Barrett.
\newblock {Efficient Neural Network Analysis with Sum-of-Infeasibilities}.
\newblock In {\em Proc. 28th Int. Conf. on Tools and Algorithms for the
  Construction and Analysis of Systems (TACAS)}, pages 143--163, 2022.

\bibitem{wu2015deep}
R.~Wu, S.~Yan, Y.~Shan, Q.~Dang, and G.~Sun.
\newblock {Deep Image: Scaling up Image Recognition}.
\newblock Technical Report. \url{http://arxiv.org/abs/1501.02876}.

\bibitem{XiTrJo18}
W.~Xiang, H.~Tran, and T.~Johnson.
\newblock {Output Reachable Set Estimation and Verification for Multi-Layer
  Neural Networks}.
\newblock {\em IEEE Transactions on Neural Networks and Learning Systems
  (TNNLS)}, 2018.

\bibitem{YaYaTrHoJoPo21}
X.~Yang, T.~Yamaguchi, H.-D. Tran, B.~Hoxha, T.~Johnson, and D.~Prokhorov.
\newblock {Neural Network Repair with Reachability Analysis}, 2021.
\newblock Technical Report. \url{https://arxiv.org/abs/2108.04214}.

\bibitem{ZhShGuGuLeNa20}
H.~Zhang, M.~Shinn, A.~Gupta, A.~Gurfinkel, N.~Le, and N.~Narodytska.
\newblock {Verification of Recurrent Neural Networks for Cognitive Tasks via
  Reachability Analysis}.
\newblock In {\em Proc. 24th European Conf. on Artificial Intelligence (ECAI)},
  pages 1690--1697, 2020.

\end{thebibliography}




\end{document}